\documentclass[floats,twocolumn,pra]{revtex4}%
\usepackage{graphics}
\usepackage{amsmath}
\usepackage{graphicx}
\usepackage{amsfonts}
\usepackage{amssymb}
\usepackage{eurosym}
\usepackage{graphics}
\usepackage{float}
\usepackage{longtable}
\usepackage{epsfig}
\usepackage{latexsym}
\usepackage{theorem}
\usepackage{bbm}
\usepackage{bm}
\usepackage{psfrag}
\usepackage{color}%
\providecommand{\U}[1]{\protect\rule{.1in}{.1in}}
\providecommand{\U}[1]{\protect\rule{.1in}{.1in}}

\begin{document}
\title{Composable security for continuous variable quantum key distribution: \\Trust levels and practical key rates in wired and wireless networks}
\author{Stefano Pirandola}
\affiliation{Department of Computer Science, University of York, York YO10 5GH, United Kingdom}

\begin{abstract}
Continuous variable (CV) quantum key distribution (QKD) provides a powerful
setting for secure quantum communications, thanks to the use of
room-temperature off-the-shelf optical devices and the potential to reach much
higher rates than the standard discrete-variable counterpart. In this work, we
provide a general framework for studying the composable finite-size security
of CV-QKD with Gaussian-modulated coherent-state protocols under various
levels of trust for the loss and noise experienced by the parties. Our study
considers both wired (i.e., fiber-based) and wireless (i.e., free-space)
quantum communications. In the latter case, we show that high key rates are
achievable for short-range optical wireless (LiFi) in secure quantum networks
with both fixed and mobile devices. Finally, we extend our investigation to
microwave wireless (WiFi) discussing security and feasibility of CV-QKD for
very short-range applications.

\end{abstract}
\maketitle

%\title{Practical security and key rates for continuous variable quantum key distribution:
%\\ Analysis of security levels for wired and wireless quantum communcations}

\section{Introduction}

Quantum key distribution (QKD)~\cite{QKDreview} enables the generation of
secret keys between two or more authenticated parties by resorting to the
fundamental laws of quantum mechanics. Its continuous variable (CV)
version~\cite{Cerf,GG02, noswitch, CVMDI,RMP} represents a very profitable
setting and opportunity thanks to its more direct implementation in the
current communication infrastructure and, most importantly, for its potential
to approach the ultimate rate limits of quantum communication, as represented
by the repeaterless PLOB bound~\cite{QKDpaper}. From an experimental point of
view, we have been witnessing an increasing number of realizations closing the
gap with the more traditional qubit-based
implementations~\cite{LeoCodes,LeoEXP}.

The most advanced protocols of CV-QKD are the Gaussian-modulated
coherent-state protocols~\cite{GG02, noswitch, CVMDI}. Not only they are very
practical, but also enjoy the most advanced security proofs, accounting for
finite-size effects (i.e., finite number of signal exchanges) and
composability (so that each step of the protocol has an associated error which
adds to an overall `epsilon'-security)~\cite{QKDreview,noteDalpha}. Very
recently, this level of security has been extended to the free-space
setting~\cite{FSpaper,SATpaper}, where we need to consider not only the
presence of diffraction-induced loss~\cite{Goodman,Siegman,svelto},
atmospheric extinction~\cite{Huffman} and background thermal noise~\cite{Miao,
BrussSAT}, but also the effect of fading, as induced by pointing error and
turbulence~\cite{Vasy12,Esposito,Yura73, Fante75,AndrewsBook, Majumdar,
Hemani}. The importance of studying fading and atmospheric effects in CV-QKD
is an active area with increasing efforts put by the community at large (e.g.
see
Refs.~\cite{refC1,refC1b,refC2,refC2b,PanosFading,refC4,refC5,refC6,refC7,refC8,refC9,refC10}%
).

While composable security is typically assessed against collective or coherent
attacks, experiments may involve some additional (realistic) assumptions that
elude this theory. For instance, these assumptions may concern some level of
trusted noise in the setups (e.g., this is often the case for the electronic
noise of the detector) or some realistic constraint on the eavesdropper, Eve
(e.g., it may be considered to be passive in line-of-sight free-space
implementations). For this reason, here we present the general theory to cover
all these cases.

In fact, we consider various levels of trust for the receiver's setup,
starting from the traditional scenario where detector's loss or noise are
untrusted, meaning that Eve may perform a side-channel attack over the
receiver besides attacking the main channel. Then, we consider the case where
detector's noise is trusted but not its loss, which corresponds to Eve
collecting leakage from the receiver. Finally, we study the more trustful
scenario where both detector's loss and noise are considered to be trusted, so
that Eve is excluded from side-channels to the receiver. We show how these
assumptions can non-trivially increase the composable key rates of
Gaussian-modulated CV-QKD protocols and tolerate higher dBs.

In our analysis, we then investigate the free-space setting, specifically for
near-range wireless quantum communications at optical frequencies (LiFi). This
scenario involves the presence of free-space diffraction and also fading
effects, mainly due to pointing and tracking errors associated with the
limited technology of the transmitter (while we can neglect turbulence at such
distances). We consider communication with both fixed and mobile devices,
assuming realistic parameters for indoor conditions and relatively-large
field-of-views for the receivers. Security is studied under the various
trusted models for the receiver's detector and then including additional
assumptions for Eve due to the line-of-sight configuration. Here too we show
that key rates are remarkably increased as an effect of the realistic
assumptions. More interestingly, we show that wireless high-rate CV-QKD is
indeed feasible with mobile devices.

Finally, we consider wireless quantum communications at the microwave
frequencies (WiFi) where both loss and thermal noise are very high. In this
scenario, we consider a potential regime of parameters that enables very
short-range quantum security, e.g., between contact-less devices within the
range of a few centimeters.

The paper is organized as follows. In Sec.~\ref{SEC1}, we provide a general
framework for the composable security of CV-QKD, which also accounts for
levels of trust in the loss and noise of the communication. In Sec.~\ref{SEC2}%
, we consider near-range free-space quantum communications, first at optical
frequencies (with fixed and mobile devices) and then at the microwaves.
Sec.~\ref{SEC3} is for conclusions.

\section{General framework for composable security of CV-QKD\label{SEC1}}

\subsection{General description\label{descriptionPROTsec}}

Let us consider a Gaussian-modulated coherent-state protocol between Alice
(transmitter) and Bob (receiver). Alice prepares a coherent state $\left\vert
\alpha\right\rangle $ whose amplitude $\alpha$ is modulated according to a
complex Gaussian distribution with zero mean and variance $\mu-1$. Assuming
the notation of Ref.~\cite{RMP}, we may decompose the amplitude as
$\alpha=(q+ip)/2$, where $x=q$ or $p$ represents the mean value of the generic
quadrature operator $\hat{x}=\hat{q},\hat{p}$ where $[\hat{q},\hat{p}]=2i$.
This generic quadrature can be written as $\hat{x}=\hat{x}_{0}+x$, where
$\hat{x}_{0}$\ is the vacuum noise associated with the bosonic mode and the
real variable $x$ is a random Gaussian displacement with zero mean and
variance
\begin{equation}
\sigma_{x}^{2}=\mu-1.
\end{equation}

The coherent state is sent through a thermal-loss channel controlled by the
eavesdropper, with transmissivity $\eta_{\text{ch}}$ and mean number of
thermal photons $\bar{n}_{e}$. Equivalently, we may introduce the variance
$\omega=2\bar{n}_{e}+1$ and the background thermal noise $\bar{n}_{B}$ defined
by $\bar{n}_{e}=\bar{n}_{B}/(1-\eta_{\text{ch}})$, so $\bar{n}_{B}$ photons
are added to the input signal. Bob's setup is characterized by quantum
efficiency $\eta_{\text{eff}}$ and extra noise variance $\nu_{\text{ex}}%
=2\bar{n}_{\text{ex}}$, where $\bar{n}_{\text{ex}}$ is an equivalent number of
thermal photons generated by the imperfections in his receiver station (due to
electronic noise, phase errors etc.)

\begin{figure}[t]
\vspace{-0.9cm}
\par
\begin{center}
\includegraphics[width=1\columnwidth] {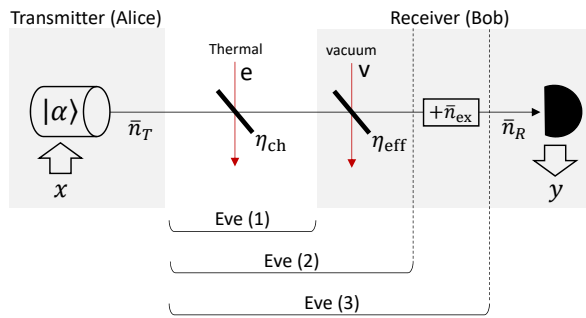}
\end{center}
\par
\vspace{-1.2cm}\caption{Quantum communication scenario between transmitter
(Alice) and receiver (Bob) separated by a quantum channel with transmissivity
$\eta_{\text{ch}}$ and thermal number $\bar{n}_{e}=\bar{n}_{B}/(1-\eta
_{\text{ch}})$.\ Bob's setup has quantum efficiency $\eta_{\text{eff}}$\ and
extra thermal photons $\bar{n}_{\text{ex}}$. The mean number of photons at the
input ($\bar{n}_{T}$) and output ($\bar{n}_{R}$) follow Eq.~(\ref{IOenergy}),
while the input classical variable ($x$)\ and the output one ($y$) follow
Eq.~(\ref{IOnew}). We also describe the various trust levels for the receiver.
In the scenario \textquotedblleft Eve~(1)\textquotedblright, the eavesdropper
is assumed to attack the external channel only. In the scenario
\textquotedblleft Eve~(2)', there is also a passive side-channel attack where
the eavesdropper collects leakage from the receiver's setup. Finally, in the
scenario \textquotedblleft Eve~(3)\textquotedblright, we assume that the
eavesdropper is also able to perform an active side-channel attack, so that
the noise internal to the setup has to be considered untrusted.}%
\label{attackPIC}%
\end{figure}

From an energetic point of view, the initial mean photons at the transmitter
$\bar{n}_{T}$ are attenuated by an overall factor $\tau=\eta_{\text{ch}}%
\eta_{\text{eff}}$ which can be seen as the total effective transmissivity of
the extended channel between Alice and Bob. Thus, the total mean number of
photons that are seen by the receiver's detector is given by%
\begin{equation}
\bar{n}_{R}=\tau\bar{n}_{T}+\bar{n}, \label{IOenergy}%
\end{equation}
where $\bar{n}$ is the total number of thermal photons due to the various
sources of noise, given by
\begin{equation}
\bar{n}=\eta_{\text{eff}}\bar{n}_{B}+\bar{n}_{\text{ex}}. \label{nBARvalue}%
\end{equation}
See also Fig.~\ref{attackPIC} for a schematic of the overall scenario.

Bob's detection is either a randomly-switched homodyne, measuring $\hat{q}$ or
$\hat{p}$ \cite{GG02}, or heterodyne, realizing the joint measurement of
$\hat{q}$ and $\hat{p}$~\cite{noswitch}. We may treat both cases compactly
with the same formalism. In both protocols, Bob retrieves an outcome $y$ which
corresponds to Alice's input $x$. For the homodyne protocol, there is a single
pair $(x,y)$ for each mode transmitted by Alice while, for the heterodyne
protocol, there are two pairs of variables per mode (but affected by more noise).

The input-output relation for the total channel from the classical input $x$
to the output $y$ takes the form
\begin{equation}
y=\sqrt{\tau}x+z, \label{IOnew}%
\end{equation}
where $z$ is a noise variable. The latter is given by
\begin{align}
z  &  =\sqrt{\eta_{\text{eff}}(1-\eta_{\text{ch}})}\hat{x}_{e}+\sqrt{\tau}%
\hat{x}_{0}\nonumber\\
&  +\sqrt{1-\eta_{\text{eff}}}\hat{x}_{v}+z_{\text{ex}}+z_{\text{det}},
\end{align}
where $\hat{x}_{e}$ denotes the quadrature of the thermal mode $e$, $\hat
{x}_{v}$ is the quadrature associated with setup vacuum mode $v$ (quantum
efficiency), $z_{\text{ex}}$ is a Gaussian variable with $\mathrm{var}%
(z_{\text{ex}})=2\bar{n}_{\text{ex}}$ accounting for the extra noise of the
setup,\ and $z_{\text{det}}$ is an additional Gaussian variable with
$\mathrm{var}(z_{\text{det}})=\nu_{\text{det}}-1$ where $\nu_{\text{det}}$ is
the quantum duty (`qu-duty') associated with detection: $\nu_{\text{det}}=1$
for homodyne and $\nu_{\text{det}}=2$ for heterodyne. See also
Fig.~\ref{attackPIC}. In total the noise variable $z$ has variance%
\begin{equation}
\sigma_{z}^{2}=2\bar{n}+\nu_{\text{det}}. \label{sigmazed}%
\end{equation}

From the input-output relation of Eq.~(\ref{IOnew}), we may compute Alice and
Bob's mutual information $I(x:y)$ which takes the same expression in direct
reconciliation (where Bob infers $x$ from $y$) and reverse reconciliation
(where Alice infers $y$ from $x$). In fact, from $\mathrm{var}(y)=\tau
\sigma_{x}^{2}+\sigma_{z}^{2}$ and $\mathrm{var}(y|x)=\sigma_{z}^{2}$, we get%
\begin{equation}
I(x:y)=\frac{\nu_{\text{det}}}{2}\log_{2}\left(  1+\frac{\sigma_{x}^{2}}{\chi
}\right)  , \label{MutualINFOeq}%
\end{equation}
where
\begin{equation}
\chi:=\frac{\sigma_{z}^{2}}{\tau}=\frac{2\bar{n}+\nu_{\text{det}}}{\tau}%
\end{equation}
is the equivalent noise. Clearly $I(x:y)$ can be specified to $I^{\text{hom}}$
(for homodyne)\ and $I^{\text{het}}$ (for heterodyne) by choosing the
corresponding value for $\nu_{\text{det}}$.

Note that the equivalent noise can be re-written as
\begin{equation}
\chi=\xi_{\text{tot}}+\frac{\nu_{\text{det}}}{\tau},~\xi_{\text{tot}}%
:=\frac{2\bar{n}}{\tau}, \label{totexcess}%
\end{equation}
where $\xi_{\text{tot}}$ defines the total excess noise. In turn, the total
excess noise can be decomposed as%
\begin{align}
\xi_{\text{tot}}  &  =\xi_{\text{ch}}+\xi_{\text{ex}},\\
\xi_{\text{ch}}  &  :=\frac{2(\bar{n}-\bar{n}_{\text{ex}})}{\tau}=\frac
{2\eta_{\text{eff}}\bar{n}_{B}}{\tau},\label{chexcess}\\
\xi_{\text{ex}}  &  :=\frac{2\bar{n}_{\text{ex}}}{\tau}, \label{excess}%
\end{align}
where $\xi_{\text{ch}}$ is the excess noise of the external channel, i.e.,
related to the thermal background, while $\xi_{\text{ex}}$ is that associated
with the extra noise in the setup.

Let us make an important remark on notation. The use of the excess noise
$\xi_{\text{tot}}$ is typical in fiber-based communication channels, while the
use of the equivalent number of thermal photons $\bar{n}$ is instead more
appropriate for free-space channels. In general, the two notations are related
by the formulas above and can be used interchangeably. In the following, we
choose to work with $\bar{n}$ which is particularly convenient from the point
of view of the finite-size estimators. However, for completeness, we also
provide the corresponding formulations in terms of excess noise.

\subsection{Local oscillator and setup noise}

Before discussing security aspects, let us discuss the local oscillator (LO)
and then clarify the main contributions to the setup noise. In terms of
equivalent number of thermal photons, the setup noise can be decomposed as
$\bar{n}_{\text{ex}}=\bar{n}_{\text{LO}}+\bar{n}_{\text{el}}+\bar
{n}_{\text{other}}$, where $\bar{n}_{\text{LO}}$ is the mean number of thermal
photons associated with the phase errors of the LO, $\bar{n}_{\text{el}}$ is
the mean number of thermal photons generated by electronic noise, and $\bar
{n}_{\text{other}}$ is any other uncharacterized but independent source of
noise (here neglected). Similarly, we may write a corresponding decomposition
in terms of excess noise $\xi_{\text{ex}}=\xi_{\text{LO}}+\xi_{\text{el}}%
+\xi_{\text{other}}$, which is obtained by using $\xi_{(...)}=2\bar{n}%
_{(...)}/\tau$.

\subsubsection{Phase-locking via TLO\ or phase-reconstruction via
LLO\label{LLOTLO}}

LO\ is crucial in CV-QKD\ since it contains the phase information that allows
the parties to exploit the two quadratures of the mode. In other words,
Alice's and Bob's rotating reference frames need to be phase-locked so Bob can
measure the incoming state in the same quadrature(s) chosen by Alice. To
achieve this goal there are two techniques, the simplest solution of the
transmitted LO (TLO)~\cite{GG02} and the more challenging (but more secure)
one of the local LO\ (LLO)~\cite{LLO,LLO2,LLO3,QKDreview}.

With the TLO, the LO is generated by the transmitter and multiplexed in
polarization with the signal mode/pulse. Both of them are sent through the
channel and then de-multiplexed by the receiver before being interfered in the
homodyne/heterodyne setup. With the LLO, bright reference pulses are regularly
interleaved with the signal pulses (time multiplexing). At the receiver, both
the signals and the references are measured with an independent local LO. From
the references, Bob is able to track Alice's rotating frame and, using this
phase information, he suitably rotates the outcomes obtained from the signals
in the phase space.

Note that both TLO and LLO\ require to employ half of the total pulses for
phase locking or reconstruction. When we explicitly consider a clock $C$ for
the system (pulses per second), the LLO involves an extra factor $1/2$ in
front of the final key rate, unless this is compensated by using both the
polarizations for the signal transmissions (not possible for the TLO).

\subsubsection{Contributions to setup noise}

From the point of view of the setup noise, we need to account for phase errors
introduced by an imperfect LO. In TLO this is negligible ($\bar{n}%
_{\text{TLO}}\simeq0$), while for the LLO it is non-trivial. In fact, assume
that signal and reference pulses are generated with an average linewidth
$l_{\text{W}}=(l_{\text{W}}^{\text{signal}}+l_{\text{W}}^{\text{LO}})/2$.
Then, for input classical modulation $\sigma_{x}^{2}$ and transmissivity
$\tau$, we may write~\cite{FSpaper}
\begin{equation}
\bar{n}_{\text{LLO}}\simeq\Theta_{\text{ph}}\tau,~\Theta_{\text{ph}}%
:=\pi\sigma_{x}^{2}C^{-1}l_{\text{W}}. \label{UBnex}%
\end{equation}
This contribution can equivalently be written as excess noise $\xi
_{\text{LLO}}=2\bar{n}_{\text{LLO}}/\tau$, according to Eq.~(\ref{excess}).
For a cw-laser $l_{\text{W}}\simeq1.6$~KHz, a clock $C=5~$MHz and a typical
modulation $\sigma_{x}^{2}=9$ (i.e., $\mu=10$) one has $\xi_{\text{LLO}}%
\simeq0.018$.

While the LLO introduces phase errors, it may actually be better when we
consider the impact of electronic noise. The latter can be described by a
variance $\nu_{\text{el}}$ or an equivalent number of photons $\bar
{n}_{\text{el}}=\nu_{\text{el}}/2$. Its value depends on the frequency of the
light $\nu$, features of the homodyne/heterodyne detector, such as its
noise equivalent power (NEP) and the bandwidth $W$, as well as features of the
LO, such as its power at detection $P_{\text{LO}}^{\text{det}}$\ and the
duration of its pulses $\Delta t_{\text{LO}}$. In fact, we may write
\begin{equation}
\bar{n}_{\text{el}}=\frac{\nu_{\text{det}}\mathrm{NEP}^{2}W\Delta
t_{\text{LO}}}{2h\nu P_{\text{LO}}^{\text{det}}}.
\end{equation}

In the case of a TLO, one has $P_{\text{LO}}^{\text{det}}=\tau P_{\text{LO}}$,
where $P_{\text{LO}}$ is the LO\ initial power at the transmitter. For an LLO,
we instead have $P_{\text{LO}}^{\text{det}}=P_{\text{LO}}$. Thus, by setting
\begin{equation}
\Theta_{\text{el}}:=\frac{\nu_{\text{det}}\mathrm{NEP}^{2}W\Delta
t_{\text{LO}}}{2h\nu P_{\text{LO}}}, \label{eleCONTRIBUT}%
\end{equation}
we may write%
\begin{equation}
\bar{n}_{\text{el}}^{\text{TLO}}=\frac{\Theta_{\text{el}}}{\tau},~\bar
{n}_{\text{el}}^{\text{LLO}}=\Theta_{\text{el}},
\end{equation}
so the formulas for the total setup noise are
\begin{equation}
\bar{n}_{\text{ex}}^{\text{TLO}}\simeq\frac{\Theta_{\text{el}}}{\tau},~\bar
{n}_{\text{ex}}^{\text{LLO}}\simeq\Theta_{\text{el}}+\Theta_{\text{ph}}\tau.
\label{setupNoiseExp}%
\end{equation}
These formulas are in terms of equivalent number of thermal photons and they
have corresponding expressions in terms of setup excess noise by using
$\xi_{\text{ex}}=2\bar{n}_{\text{ex}}/\tau$.

\begin{figure}[t]
\vspace{0.2cm}
\par
\begin{center}
\includegraphics[width=0.70\columnwidth] {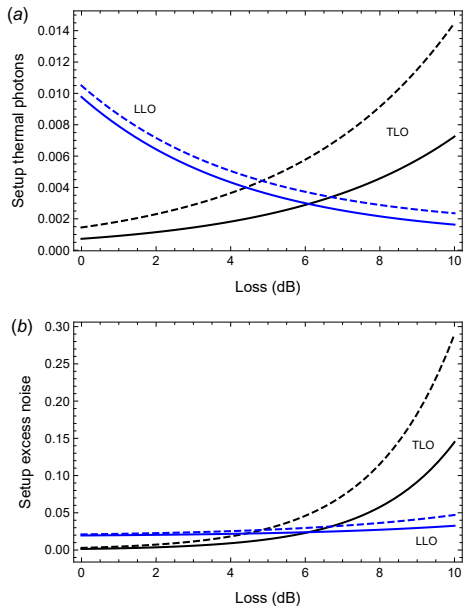}
\end{center}
\par
\vspace{-0.5cm}\caption{Setup noise as a function of the total transmissivity
$\tau$ expressed in decibels. (a)~We plot the equivalent number of thermal
photons $\bar{n}_{\text{ex}}$ associated with the setup noise, for the TLO
(black lines) and the LLO\ (blue lines), considering the homodyne protocol
(solid lines) and the heterodyne protocol (dashed lines). (b)~As in~(a) but we
plot the setup excess noise $\xi_{\text{ex}}$. Parameters are chosen as in the
text. See Eq.~(\ref{setupNoiseExp}).}%
\label{TLOLLO}%
\end{figure}

Above we can see the different monotonicity of the setup noise with respect to
$\tau$, between TLO\ and LLO. Assume $\lambda=800~$nm and $W=100~$MHz, so we
have signal pulses of duration $\Delta t=10~$ns and LO pulses of duration
$\Delta t_{\text{LO}}=10~$ns. For this bandwidth, we can assume the good value
$\mathrm{NEP}=6~$pW/$\sqrt{\text{Hz}}$. Then, assuming $P_{\text{LO}}=100~$mW,
we get $\Theta_{\text{el}}\simeq1.45\times10^{-3}$ for heterodyne detection
($\nu_{\text{det}}=2$). For the LLO this value remains low, while for the TLO
it is rescaled by $1/\tau$, which means that it may become large at long
distances. See also Fig.~\ref{TLOLLO}\ for a comparison.

\subsection{Trust levels\label{Section_levels}}

Once we have clarified the main sources of noise in the communication
scenario, we can go ahead and identify different levels of trust on the basis
of different assumptions for the eavesdropper (Eve). The basic model is to
assume that Eve's action is restricted to the outside channel. In this
strategy, she inserts her photons in the thermal background and stores all the
photons which are not collected by the receiver. However, she is assumed not
to monitor or control the receiver's setup. This is the scenario where loss
and noise are considered to be trusted in the receiver. See also Eve~(1) in
Fig.~\ref{attackPIC}. In this case, Eve's collective Gaussian attack is
represented by a purification of the environmental beam-splitter of
transmissivity $\eta_{\text{ch}}$, where the injected $\bar{n}_{e}^{(1)}%
=\bar{n}_{B}(1-\eta_{\text{ch}})^{-1}$ thermal photons are to be considered
part of a two-mode squeezed vacuum (TMSV) state in Eve's
hands~\cite{collectiveG}.

More generally, we can assume that Eve is able to detect the leakage from
setups~\cite{refB1,refB2,refB3}. Here we consider this potential problem for
the receiver's setup, so that the fraction $1-\eta_{\text{eff}}$ of the
photons missed by the detection is stored by Eve and becomes part of her
attack. On the other hand, we may assume that Eve is not able to actively
tamper with the receiver, i.e., she does not control the noise internal to the
setup, which may therefore be considered as trusted (this is a reasonable
assumption which is often made by experimentalists for the electronic noise of
the detector). We call this scenario the trusted-noise model for the receiver.
See Eve~(2) in Fig.~\ref{attackPIC}. In this case, the efficiency
$\eta_{\text{eff}}$ becomes part of Eve's environmental beam-splitter, which
now has total transmissivity $\tau=\eta_{\text{ch}}\eta_{\text{eff}}$ and
injects $\bar{n}_{e}^{(2)}=\eta_{\text{eff}}\bar{n}_{B}(1-\tau)^{-1}$ thermal photons.

Finally, there is the worst-case scenario where no imperfection in the
receiver setup is trusted. In fact, the most pessimistic assumption is that
Eve can also potentially control the extra photons in the setup $\bar
{n}_{\text{ex}}$ besides collecting its leakage. See also Eve~(3) in
Fig.~\ref{attackPIC}. In this case, the extra photons become part of Eve's
environment. In other words, the entire channel from the transmitter to the
final (ideal) detection is dilated into a single beam-splitter with
transmissivity $\tau=\eta_{\text{ch}}\eta_{\text{eff}}$ and injecting $\bar
{n}_{e}^{(3)}=\bar{n}(1-\tau)^{-1}$ thermal photons.

Clearly the security increases from the completely trusted receiver [Eve~(1)]
to the worst-case scenario [Eve~(3)]. Similarly, the key rate will decrease,
because more degrees of freedom would go under Eve's control. For this reason,
the worst-case scenario provides a lower bound for all the others. Also note
that the worst-case scenario progressively collapses in the lower levels if we
assume $\bar{n}_{\text{ex}}=0$ and then $\eta_{\text{eff}}=1$. Also note that,
in general, one \ may consider hybrid situations between Eve~(2) and Eve~(3),
where the setup noise $\bar{n}_{\text{ex}}$ is partly trusted ($\bar
{n}_{\text{ex}}^{\text{tr}}$) and partly untrusted ($\bar{n}_{\text{ex}%
}^{\text{unt}}$). This is included by writing $\bar{n}_{\text{ex}}%
^{\text{unt}}=\eta_{\text{eff}}\bar{n}_{B}^{\text{unt}}$ and increasing the
background $\bar{n}_{B}\rightarrow\bar{n}_{B}+\bar{n}_{B}^{\text{unt}}$.

\subsection{Asymptotic key rates\label{subsectionAKR}}

It is convenient to start by studying the security of the protocol with the
intermediate assumption of a trusted-noise detector as in
Fig.~\ref{dilationPIC2}, where the setup noise is considered to be trusted,
i.e., not coming from Eve's attack [cf. Eve~(2) in Fig.~\ref{attackPIC}].
Then, we analyze the key rate in the most optimistic case where also the setup
loss is considered to be trusted. Finally, we compare the formulas with the
worst-case scenario, where all noise is considered to be untrusted [cf.
Eve~(3) in Fig.~\ref{attackPIC}]. The latter represents the case analyzed in
Ref.~\cite{FSpaper}.

\subsubsection{Asymptotic key rate with a trusted-noise detector}

\begin{figure}[t]
\vspace{-1.7cm}
\par
\begin{center}
\includegraphics[width=0.5\textwidth] {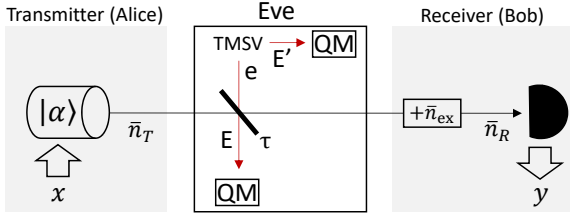}
\end{center}
\par
\vspace{-2.0cm}\caption{Eve's collective attack under the assumption of
trusted noise in the receiver's setup, i.e., Eve~(2) in Fig.~\ref{attackPIC}.}%
\label{dilationPIC2}%
\end{figure}

Consider the trusted-noise detector which corresponds to the dilated scenario
in Fig.~\ref{dilationPIC2}. Here the total transmissivity is $\tau
=\eta_{\text{ch}}\eta_{\text{eff}}$ and the injected thermal noise is given by
$\bar{n}_{e}^{(2)}=\eta_{\text{eff}}\bar{n}_{B}(1-\tau)^{-1}$. In order to
compute the asymptotic secret key rate in reverse reconciliation, we consider
Bob and Eve's joint covariance matrix (CM). Let us define the basic block
matrices $\mathbf{I}:=\mathrm{diag}(1,1)$ and $\mathbf{Z}:=\mathrm{diag}%
(1,-1)$. Then, the joint CM is given by%
\begin{equation}
\mathbf{V}_{BEE^{\prime}}=\left(
\begin{array}
[c]{cc}%
b\mathbf{I} & \mathbf{C}\\
\mathbf{C}^{T} & \mathbf{V}_{EE^{\prime}}%
\end{array}
\right)  , \label{jointCM}%
\end{equation}
where Eve's reduced CM $\mathbf{V}_{EE^{\prime}}$ and the cross-correlation
block $\mathbf{C}$ take the forms
\begin{equation}
\mathbf{V}_{EE^{\prime}}=\left(
\begin{array}
[c]{cc}%
\phi\mathbf{I} & \psi\mathbf{Z}\\
\psi\mathbf{Z} & \omega\mathbf{I}%
\end{array}
\right)  ,~\mathbf{C}=\left(
\begin{array}
[c]{cc}%
\theta\mathbf{I} & \gamma\mathbf{Z}%
\end{array}
\right)  , \label{CMcase1}%
\end{equation}
where we have set%
\begin{align}
\omega &  =2\bar{n}_{e}^{(2)}+1=\frac{2\eta_{\text{eff}}\bar{n}_{B}}{1-\tau
}+1=\frac{\tau\xi_{\text{ch}}}{1-\tau}+1,\label{EvesVAR}\\
b  &  =\tau(\mu-1)+2\bar{n}+1=\tau(\mu-1)+\tau\xi_{\text{tot}}%
+1,\label{parameterB}\\
\gamma &  =\sqrt{(1-\tau)(\omega^{2}-1)},~\theta=\sqrt{\tau(1-\tau)}%
(\omega-\mu),\label{thetaEQ1}\\
\psi &  =\sqrt{\tau(\omega^{2}-1)},~\phi=\tau\omega+(1-\tau)\mu. \label{fiEQ1}%
\end{align}

In the homodyne protocol, Eve's conditional CM\ on Bob's outcome $y$ is given
by~\cite{RMP,GaeCM,GaeCM2}%
\begin{equation}
\mathbf{V}_{EE^{\prime}|B}^{\text{hom}}=\mathbf{V}_{EE^{\prime}}%
-b^{-1}\mathbf{C}^{T}\boldsymbol{\Pi}\mathbf{C},
\end{equation}
where $\boldsymbol{\Pi}:=\mathrm{diag}(1,0)$. In the heterodyne protocol, we
have instead the following conditional CM~\cite{RMP,GaeCM,GaeCM2}%
\begin{equation}
\mathbf{V}_{EE^{\prime}|B}^{\text{het}}=\mathbf{V}_{EE^{\prime}}%
-(b+1)^{-1}\mathbf{C}^{T}\mathbf{C}.
\end{equation}

Call $\{\nu_{\pm}\}$ the symplectic spectrum of Eve's CM\ $\mathbf{V}%
_{EE^{\prime}}$. Then, call $\{\nu_{\pm}^{\text{hom}}\}$ and $\{\nu_{\pm
}^{\text{het}}\}$ the symplectic spectra of Eve's conditional CMs
$\mathbf{V}_{EE^{\prime}|B}^{\text{hom}}$ and $\mathbf{V}_{EE^{\prime}%
|B}^{\text{het}}$, respectively. Then, we may compute Eve's Holevo information
for both protocols, as%
\begin{align}
\chi^{\text{hom}}(\mathbf{E}  &  :y)=\sum\limits_{k=\pm}\left[  H(\nu
_{k})-H(\nu_{k}^{\text{hom}})\right]  ,\label{homoCHI}\\
\chi^{\text{het}}(\mathbf{E}  &  :y)=\sum\limits_{k=\pm}\left[  H(\nu
_{k})-H(\nu_{k}^{\text{het}})\right]  , \label{heteroCHI}%
\end{align}
where $\mathbf{E}=EE^{\prime}$ and $H(x)$ is the entropic function%
\begin{equation}
H(x):=\frac{x+1}{2}\log_{2}\frac{x+1}{2}-\frac{x-1}{2}\log_{2}\frac{x-1}{2}.
\end{equation}

For a realistic reconciliation efficiency $\beta\in\lbrack0,1]$, accounting
for the fact that data-processing may not reach the Shannon limit, we write
the asymptotic key rate
\begin{equation}
R_{\text{asy}}^{(2)}(\tau,\bar{n},\bar{n}_{B})=\beta I(x:y)_{\tau,\bar{n}%
}-\chi(\mathbf{E}:y)_{\tau,\bar{n},\bar{n}_{B}}, \label{rateASYm}%
\end{equation}
where the explicit expressions for the homodyne protocol~\cite{GG02} and the
heterodyne protocol~\cite{noswitch} derive from the corresponding expressions
for the mutual information [cf. Eq.~(\ref{MutualINFOeq})] and the Holevo bound
[cf. Eqs.~(\ref{homoCHI}) and~(\ref{heteroCHI})].

It is clear that, in a practical setting, the parties do not know all the
parameters entering the rate in Eq.~(\ref{rateASYm}), so they need to resort
to suitable procedures of parameter estimation. It is acceptable to assume
that Alice controls/knows the signal modulation $\mu$, while Bob
monitors/knows the quantum efficiency $\eta_{\text{eff}}$. The channel
parameters $\tau$ and $\bar{n}$ need to be estimated. In general, the setup noise
$\bar{n}_{\text{ex}}$ depends on the total transmissivity $\tau$. For this
reason, $\bar{n}_{\text{ex}}$ too needs to be estimated by the parties. The
estimates of $\bar{n}$ and $\bar{n}_{\text{ex}}$ then provide the value of
$\bar{n}_{B}$.

\subsubsection{Asymptotic key rate with a trusted-loss and trusted-noise
detector}

Here we consider the best possible scenario for Alice and Bob, which is the
assumption of Eve~(1) in Fig.~\ref{attackPIC}. Not only the setup noise is
trusted but also the loss of the setup into the external environment is
considered to be trusted (i.e., we assume Eve is not collecting the leakage
from the setup). The asymptotic key rate can be found by a simple modification
of the previous derivation.

From the point of view of Alice and Bob, the mutual information is clearly the
same. For Eve instead, the effective beam splitter used in her attack has now
transmissivity $\eta_{\text{ch}}$ and input thermal noise $\bar{n}_{e}%
^{(1)}=\bar{n}_{B}(1-\eta_{\text{ch}})^{-1}$. It is easy to check that we need
to use the CM in Eq.~(\ref{CMcase1}) with the replacements%
\begin{align}
\omega &  =2\bar{n}_{e}^{(1)}+1=\frac{2\bar{n}_{B}}{1-\eta_{\text{ch}}%
}+1=\frac{\eta_{\text{ch}}\xi_{\text{ch}}}{1-\eta_{\text{ch}}}+1,\\
\gamma &  =\sqrt{\eta_{\text{eff}}(1-\eta_{\text{ch}})(\omega^{2}-1)},\\
\theta &  =\sqrt{\tau(1-\eta_{\text{ch}})}(\omega-\mu),\label{thetaEQ2}\\
\psi &  =\sqrt{\eta_{\text{ch}}(\omega^{2}-1)},~\phi=\eta_{\text{ch}}%
\omega+(1-\eta_{\text{ch}})\mu, \label{fiEQ2}
\end{align}
while parameter $b$ is the same as in Eq.~(\ref{parameterB}).

The next steps are as before. One computes the symplectic spectrum $\{\nu
_{\pm}\}$ of the CM $\mathbf{V}_{EE^{\prime}}$ and those, $\{\nu_{\pm
}^{\text{hom}}\}$ and $\{\nu_{\pm}^{\text{het}}\}$, of the conditional CMs
$\mathbf{V}_{EE^{\prime}|B}^{\text{hom}}$ and $\mathbf{V}_{EE^{\prime}%
|B}^{\text{het}}$. These eigenvalues are then replaced in Eqs.~(\ref{homoCHI})
and~(\ref{heteroCHI}). In this way, we get the corresponding asymptotic key
rates $R_{\text{asy}}^{(1)}(\tau,\bar{n},\bar{n}_{B})$ following the formula
in Eq.~(\ref{rateASYm}). Parameters need to be estimated in the same way as
explained in the previous subsection.

\subsubsection{Asymptotic key rate with untrusted
detector\label{AsyUntrustedSection}}

In the worst-case scenario of untrusted noise [cf. Eve~(3) in
Fig.~\ref{attackPIC}], the entire channel is dilated into a single beam
splitter with transmissivity $\tau=\eta_{\text{ch}}\eta_{\text{eff}}$, where
Eve injects $\bar{n}_{e}^{(3)}=\bar{n}(1-\tau)^{-1}$ thermal photons. Setup
noise $\bar{n}_{\text{ex}}$ becomes part of Eve's attack, so all excess noise
is now considered to be untrusted. From the point of view of the asymptotic
key rate, it is sufficient to replace $\eta_{\text{eff}}\bar{n}_{B}=\bar
{n}-\bar{n}_{\text{ex}}\rightarrow\bar{n}$ in the expression of Eve's variance
$\omega$ in Eq.~(\ref{EvesVAR}), with implicit modifications for the other
elements of the CM. More precisely, it is sufficient to set%
\begin{equation}
\omega=2\bar{n}_{e}^{(3)}+1=\frac{2\bar{n}}{1-\tau}+1=\frac{\tau
\xi_{\text{tot}}}{1-\tau}+1. \label{untrustedOMEGA}%
\end{equation}

Alternatively, we can exploit the entanglement-based representation of the
protocol according to which Alice's Gaussian-modulated coherent states are
realized by heterodyning mode $A$ of a TMSV state~\cite{RMP} with CM
\begin{equation}
\mathbf{V}_{AA^{\prime}}=\left(
\begin{array}
[c]{cc}%
\mu\mathbf{I} & \sqrt{\mu^{2}-1}\mathbf{Z}\\
\sqrt{\mu^{2}-1}\mathbf{Z} & \mu\mathbf{I}%
\end{array}
\right)  .
\end{equation}
After the thermal-loss channel with total transmissivity $\tau$, Alice and
Bob's shared Gaussian state $\rho_{AB}$ has CM%
\begin{equation}
\mathbf{V}_{AB}=\left(
\begin{array}
[c]{cc}%
\mu\mathbf{I} & \sqrt{\tau(\mu^{2}-1)}\mathbf{Z}\\
\sqrt{\tau(\mu^{2}-1)}\mathbf{Z} & b\mathbf{I}%
\end{array}
\right)  .
\end{equation}

Eve is assumed to hold the purification of $\rho_{AB}$, so the total state
$\rho_{AB\mathbf{E}}$ of Alice, Bob and Eve is pure. This means that
$S(\mathbf{E})=S(AB)$, where $S(Q)$ denotes the von Neumann entropy computed
over the state $\rho_{Q}$ of system $Q$. Then, because homodyne/heterodyne is
a rank-1 measurement (projecting pure states in pure states), we have that
$\rho_{A\mathbf{E}|y}$ is pure, which implies the equality of the conditional
entropies $S(\mathbf{E}|y)=S(A|y)$. As a result, Eve's Holevo bound is simply
given by
\begin{equation}
\chi(\mathbf{E}:y):=S(\mathbf{E})-S(\mathbf{E}|y)=S(AB)-S(A|y).
\label{ChiJOINT}
\end{equation}

Thus, we may compute $\chi(\mathbf{E}:y)$ using Alice and Bob's CM
$\mathbf{V}_{AB}$ with symplectic eigenvalues $\nu_{\pm}^{\prime}$. It is easy
to find~\cite{FSpaper}%
\begin{align}
\chi^{\text{hom}}(\mathbf{E}  &  :y)=H(\nu_{+}^{\prime})+H(\nu_{-}^{\prime
})-H\left[  \sqrt{\mu^{2}-\frac{\mu\tau(\mu^{2}-1)}{b}}\right]  ,\\
\chi^{\text{het}}(\mathbf{E}  &  :y)=H(\nu_{+}^{\prime})+H(\nu_{-}^{\prime
})-H\left[  \mu-\frac{\tau(\mu^{2}-1)}{b+1}\right]  ,
\end{align}
where $b$ is given in Eq.~(\ref{parameterB}).

Using these expressions and the mutual information of Eq.~(\ref{MutualINFOeq}%
), we write%
\begin{equation}
R_{\text{asy}}^{(3)}(\tau,\bar{n})=\beta I(x:y)_{\tau,\bar{n}}-\chi
(\mathbf{E}:y)_{\tau,\bar{n}}. \label{asyUNrate}%
\end{equation}
Note that the parties only need to estimate the extended-channel parameters
$\tau$ and $\bar{n}$. As we see below these estimators are built up to some
error probability $\varepsilon_{\text{pe}}$.

\subsection{Parameter estimation\label{PEmainSEC}\label{PEsection}%
\label{PEsectionLAter}}

As mentioned in the previous section, Alice and Bob need to estimate some of
the parameters. Even if they control the values of the input Gaussian
modulation $\mu$ and they can calibrate the output quantum efficiency
$\eta_{\text{eff}}$, they still need to estimate the various channel's
parameters and the setup noise $\bar{n}_{\text{ex}}$. The procedure has some
differences depending if we consider a trusted or untrusted model for the
receiver. For a trusted-noise detector [Eve~(2)] and a fully-trusted detector
[Eve~(1)], Alice and Bob need to estimate $\tau$, $\bar{n}$ and $\bar{n}_{B}$
(via $\bar{n}_{\text{ex}}$). For the untrusted detector [Eve~(3)], they only
need to estimate $\tau$ and $\bar{n}$, since the two thermal contributions
$\bar{n}_{B}$ and $\bar{n}_{\text{ex}}$ are both considered to be untrusted
(and therefore merged into a single parameter).

We therefore consider two basic independent estimators $\hat{\tau}$ and
$\widehat{\bar{n}}$, for $\tau$ and $\bar{n}$. Then, in the trusted scenarios
[Eve~(1) and~(2)], we also require the use of additional estimators, which can
be derived from the basic ones. To estimate the parameters, Alice and Bob
randomly and jointly choose $m$ of the $N$ distributed signals, and publicly
disclose the corresponding $m_{p}:=\nu_{\text{det}}m$ pairs of values
$\{x_{i},y_{i}\}_{i=1}^{m_{p}}$. These are $m$ pairs for the homodyne protocol
and $2m$ pairs for the heterodyne protocol. Under the standard assumption of a
collective (entangling-cloner) Gaussian attack, these pairs are independent
and identically distributed Gaussian variables, related by Eq.~(\ref{IOnew}).

From the pairs, they build the estimator $\hat{T}$ of the square-root
transmissivity $T:=\sqrt{\tau}$, i.e.,%
\begin{equation}
\hat{T}=\frac{\sum_{i=1}^{m_{p}}x_{i}y_{i}}{\sum_{i=1}^{m_{p}}x_{i}^{2}},
\end{equation}
and the estimator $\widehat{\sigma_{z}^{2}}$ of the noise variance $\sigma
_{z}^{2}$, i.e.,
\begin{equation}
\widehat{\sigma_{z}^{2}}=\frac{1}{m_{p}}\sum_{i=1}^{m_{p}}(y_{i}-\hat{T}%
x_{i})^{2}. \label{sigmaZesimates}%
\end{equation}
From these, we can derive the two basic estimators
\begin{equation}
\hat{\tau}:=\hat{T}^{2},~\widehat{\bar{n}}:=\frac{\widehat{\sigma_{z}^{2}}%
-\nu_{\text{det}}}{2}.
\end{equation}

For a confidence parameter $w$, we then define and compute the worst-case
estimators~\cite{NoteEstimator}
\begin{align}
\tau^{\prime}  &  :=\hat{\tau}-w\sqrt{\mathrm{var}(\hat{\tau})}\simeq
\tau-2w\sqrt{\frac{2\tau^{2}+\tau\sigma_{z}^{2}/\sigma_{x}^{2}}{m_{p}}%
},\label{wcEstimator1}\\
\bar{n}^{\prime}  &  :=\widehat{\bar{n}}+w\sqrt{\mathrm{var}(\widehat{\bar{n}%
})}\simeq\bar{n}+w\frac{\sigma_{z}^{2}}{\sqrt{2m_{p}}}. \label{wcEstimator2}%
\end{align}
Each of these estimators bounds the corresponding actual value, $\tau$ and
$\bar{n}$, up to an error probability $\varepsilon_{\text{pe}}$ if we take%
\begin{equation}
w=\sqrt{2}\operatorname{erf}^{-1}(1-2\varepsilon_{\text{pe}}),
\label{wSTVALUE}%
\end{equation}
or, in case of low values ($\varepsilon_{\text{pe}}\leq10^{-17}$), if we take%
\begin{equation}
w=\sqrt{2\ln(1/\varepsilon_{\text{pe}})}. \label{wTAIL}%
\end{equation}
As a result the total error probability associated with parameter estimation
is $\simeq2\varepsilon_{\text{pe}}$. See Ref.~\cite{FSpaper} for more
technical details on these derivations, which exploit tools from
Ref.~\cite{UsenkoFinite} and involves suitable tail
bounds~\cite{TailBound,Kolar}.

For the trusted-detector scenarios, we need to provide the best-case estimator
of $\bar{n}_{\text{ex}}$, which automatically allows us to derive the
worst-case estimator of $\bar{n}_{B}$. From the analytical expressions in
Eq.~(\ref{setupNoiseExp}), we see that we need to account for the different
behavior of $\bar{n}_{\text{ex}}$ in terms of the transmissivity $\tau$, which
requires both the use of a worst-case estimator $\tau^{\prime}$ and that of a
best-case estimator $\tau^{\prime\prime}:=\hat{\tau}+w\sqrt{\mathrm{var}%
(\hat{\tau})}$. In other words, we have%
\begin{align}
\bar{n}_{\text{ex}}^{\text{TLO}}  &  \gtrsim\bar{n}_{\text{ex,bc}}%
^{\text{TLO}}:=\frac{\Theta_{\text{el}}}{\tau^{\prime\prime}},\label{bcc1}\\
\bar{n}_{\text{ex}}^{\text{LLO}}  &  \gtrsim\bar{n}_{\text{ex,bc}}%
^{\text{LLO}}:=\Theta_{\text{el}}+\Theta_{\text{ph}}\tau^{\prime}.
\label{bcc2}%
\end{align}
Correspondingly, we have the following worst-case estimator for the background
thermal noise
\begin{equation}
\bar{n}_{B}\lesssim\bar{n}_{B}^{\prime}:=\frac{\bar{n}^{\prime}-\bar
{n}_{\text{ex,bc}}}{\eta_{\text{eff}}}.
\end{equation}

We can now compute the values of the asymptotic key rates affected by
parameter estimation. For the various scenarios, these are given by
\begin{align}
R_{\text{asy}}^{(1,2)}(\tau,\bar{n},\bar{n}_{B})  &  \rightarrow\frac{n}%
{N}R_{\text{asy}}^{(1,2)}(\tau^{\prime},\bar{n}^{\prime},\bar{n}_{B}^{\prime
}),\\
R_{\text{asy}}^{(3)}(\tau,\bar{n})  &  \rightarrow\frac{n}{N}R_{\text{asy}%
}^{(3)}(\tau^{\prime},\bar{n}^{\prime}),
\end{align}
where $n=N-m$ is the number of signals left for key generation (after $m$ are
discarded for parameter estimation). These key rates are correct up to an
error $\simeq2\varepsilon_{\text{pe}}$.

As a final remark, notice that the total excess noise $\xi_{\text{tot}}$ can
be estimated by using $\hat{\tau}$ and $\widehat{\bar{n}}$\ via
Eq.~(\ref{totexcess}) and therefore worst-case estimated by using
$\tau^{\prime}$ and $\bar{n}^{\prime}$, i.e.,
\begin{equation}
\xi_{\text{tot}}\lesssim\xi_{\text{tot}}^{\prime}:=\frac{2\bar{n}^{\prime}%
}{\tau^{\prime}}.
\end{equation}
Similarly, the channel excess noise $\xi_{\text{ch}}$ can be worst-case
estimated by combining Eq.~(\ref{chexcess}) with $\tau^{\prime}$\ and $\bar
{n}_{B}^{\prime}$, i.e.,%
\begin{equation}
\xi_{\text{ch}}\lesssim\xi_{\text{ch}}^{\prime}:=\frac{2\eta_{\text{eff}}%
\bar{n}_{B}^{\prime}}{\tau^{\prime}}.
\end{equation}

\subsection{Composable finite-size key rates\label{KR_sec}}

After parameter estimation, each block of size $N$ provides $n$ signals to be
processed into a shared key via error correction and privacy amplification.
Given a block, this is successfully error-corrected with probability
$p_{\text{ec}}$ (or failure probability $\mathrm{FER}=1-p_{\text{ec}}$ known
as `frame error rate'). The value of $p_{\text{ec}}$ depends on the
signal-to-noise ratio, the target reconciliation efficiency $\beta$, and the
$\varepsilon$-correctness $\varepsilon_{\text{cor}}$, the latter bounding the
probability that Alice's and Bob's local strings are different after error
correction and successful verification of their hashes.

On average $np_{\text{ec}}$ signals per block are promoted to privacy
amplification. This final step is implemented with an associated $\varepsilon
$-secrecy $\varepsilon_{\text{sec}}$, the latter bounding the distance between
the final key and an ideal key that is completely uncorrelated from Eve. In
turn, the $\varepsilon$-secrecy is technically decomposed as $\varepsilon
_{\text{sec}}=\varepsilon_{\text{s}}+\varepsilon_{\text{h}}$, where
$\varepsilon_{\text{s}}$ is a smoothing parameter and $\varepsilon_{\text{h}}$
is a hashing parameter.

Overall, the final composable key rate of the protocol takes the
form~\cite{FSpaper}
\begin{equation}
R\geq\frac{np_{\text{ec}}}{N}\left(  R_{\text{pe}}^{(k)}-\frac{\Delta
_{\text{aep}}}{\sqrt{n}}+\frac{\Theta}{n}\right)  , \label{sckeee}%
\end{equation}
where $R_{\text{pe}}^{(k)}$ depends on the receiver model%
\begin{equation}
R_{\text{pe}}^{(1,2)}=R_{\text{asy}}^{(1,2)}(\tau^{\prime},\bar{n}^{\prime
},\bar{n}_{B}^{\prime}),~~R_{\text{pe}}^{(3)}=R_{\text{asy}}^{(3)}%
(\tau^{\prime},\bar{n}^{\prime}), \label{RpeFORMU}%
\end{equation}
and the extra finite-size terms are equal to
\begin{align}
&  \Delta_{\text{aep}}=4\log_{2}\left(  2\sqrt{d}+1\right)  \sqrt{\log
_{2}\left(  \frac{18}{p_{\text{ec}}^{2}\varepsilon_{\text{s}}^{4}}\right)
},\label{deltaAEPPP}\\
&  \Theta=\log_{2}[p_{\text{ec}}(1-\varepsilon_{\text{s}}^{2}/3)]+2\log
_{2}\sqrt{2}\varepsilon_{\text{h}}. \label{bigOMEGA}%
\end{align}
Here the parameter $d$ is the size of Alice's and Bob's effective alphabet
after analog-to-digital conversion of their continuous variables $x$ and $y$
($d=2^{5}=32$ for a $5$-bit discretization). This rate refers to security
against collective Gaussian attacks with total epsilon security~\cite{FSpaper}%
\begin{equation}
\varepsilon=2p_{\text{ec}}\varepsilon_{\text{pe}}+\varepsilon_{\text{cor}%
}+\varepsilon_{\text{sec}}. \label{epsSECcollective}%
\end{equation}

\subsubsection{Improved pre-factor}

Note that the prefactor $\log_{2}(2\sqrt{d}+1)$ in the AEP\ term in
Eq.~(\ref{deltaAEPPP}) can be tightened into $\log_{2}(\sqrt{d}+2)$. In
general, according to Theorem~6.4 and Corollary~6.5 of Ref.~\cite{TomaThesis},
one can lower-bound the conditional smooth min-entropy $H_{\min}^{\delta
}(y^{n}|\mathbf{E}^{n})$ associated with the $n$-use classical-quantum state
$\rho_{y\mathbf{E}}^{\otimes n}$\ shared between Bob (classical system $y$)
and Eve (quantum system $\mathbf{E}$). This is done by using the conditional
entropy between the single-use systems ($y$ and $\mathbf{E}$) up to a penalty,
i.e., we may write~\cite{TomaThesis,TomaBook}
\begin{equation}
H_{\min}^{\delta}(y^{n}|\mathbf{E}^{n})_{\rho^{\otimes n}}\geq nH(y|\mathbf{E}%
)_{\rho}+\sqrt{n}\Delta_{\text{aep}}(\delta),
\end{equation}
where
\begin{align}
\Delta_{\text{aep}}(\delta) &  =4(\log_{2}v)\sqrt{-\log_{2}(1-\sqrt
{1-\delta^{2}})}\nonumber\\
&  \simeq4(\log_{2}v)\sqrt{\log_{2}(2/\delta^{2})}\\
v &  \leq\sqrt{2^{-H_{\min}(y|\mathbf{E})}}+\sqrt{2^{H_{\max}(y|\mathbf{E})}%
}+1,\label{equationV}%
\end{align}
with $v$ being bounded using min- and max-entropies. Recall that the min- and
max-entropies can be negative in general, but their absolute values must be
$\leq\log_{2}d$, with $d$ being the size of Bob's alphabet (e.g., this easily
follows from Ref.~\cite[Lemma~5.2]{TomaBook}). This implies the bound
$v\leq2\sqrt{d}+1$, which leads to the prefactor used in Eq.~(\ref{deltaAEPPP}%
). See Ref.~\cite[Appendix~G]{FSpaper} for details on how to connect the key
rate with the conditional smooth min-entropy and simplify derivations via the
AEP term.

However, it is worth noting that, for a classical-quantum state $\rho
_{y\mathbf{E}}$, the conditional min-entropy is non-negative, i.e., $H_{\min
}(y|\mathbf{E})\geq0$. This is a property that can be shown, more generally,
for separable states. In fact, starting from the definition of conditional
min-entropy for a generic state $\rho_{AB}$ of two quantum systems $A$ and $B$
\cite[Def. 4.1]{TomaThesis}, we can write the lower bound%
\begin{equation}
H_{\min}(A|B)_{\rho}\geq\tilde{H}:=\sup\{\lambda\in\mathbb{R}:\rho_{AB}%
\leq2^{-\lambda}I_{A}\otimes\rho_{B}\}.\label{eq2}%
\end{equation}
For separable $\rho_{AB}$, one may write~\cite[Lemma~5.2]{TomaBook}%
\begin{equation}
\rho_{AB}=\sum_{k}p_{k}\theta_{A}^{k}\otimes\rho_{B}^{k}\leq\sum_{k}p_{k}%
I_{A}\otimes\rho_{B}^{k}=I_{A}\otimes\rho_{B},
\end{equation}
which leads to $\tilde{H}\geq0$, since we are left to find the
\textit{maximum} value of $\lambda$ such that
\begin{equation}
\rho_{AB}\leq I_{A}\otimes\rho_{B},~\rho_{AB}\leq2^{-\lambda}I_{A}\otimes
\rho_{B}.
\end{equation}

Thus, using $H_{\min}(y|\mathbf{E})\geq0$ in Eq.~(\ref{equationV}), we may
write $v\leq\sqrt{d}+2$ which improves Eq.~(\ref{deltaAEPPP}) into
\begin{equation}
\Delta_{\text{aep}}=4\log_{2}\left(  \sqrt{d}+2\right)  \sqrt{\log_{2}\left(
\frac{18}{p_{\text{ec}}^{2}\varepsilon_{\text{s}}^{4}}\right)  }.
\end{equation}
Note that, for a typical $5$-bit digitalization $d=2^{5}$, we have $\log
_{2}(\sqrt{d}+2)\simeq2.94$ instead of $\log_{2}(2\sqrt{d}+1)\simeq3.6$, so
the improvement is limited. In our numerical investigations we assume the
worst-case pre-factor, but keeping in mind that performances can be slightly improved.

\subsubsection{Extension to coherent attacks}

For the heterodyne protocol, the key rate can be extended to security against
general attacks using tools from Ref.~\cite{Lev2017}. Let us symmetrize the
protocol by applying an identical random orthogonal matrix to the classical
continuous variables of the two parties. Then, assume that Alice and Bob
jointly perform $m_{\mathrm{et}}=f_{\mathrm{et}}n$ energy tests on randomly
chosen uses of the channel (for some factor $f_{\mathrm{et}}<1$). In each
test, the parties measure the local number of photons (which can be
extrapolated from the data) and compute an average over the $m_{\mathrm{et}}$
tests. If these averages are greater than a threshold $d_{\text{et}}$, the
protocol is aborted. Setting $d_{\text{et}}\gtrsim\bar{n}_{T}+\mathcal{O}%
(m_{\mathrm{et}}^{-1/2})$ assures secure success of the test in typical
scenarios (where signals are attenuated and noise is not too high).

The number of signals for key generation is reduced to
\begin{equation}
n=N-(m+m_{\mathrm{et}})=\frac{N-m}{1+f_{\mathrm{et}}},
\end{equation}
and the procedure needs an additional step of privacy amplification
compressing the final key by a further amount
\begin{align}
\Phi_{n}  &  :=2\left\lceil \log_{2}\binom{K_{n}+4}{4}\right\rceil
,\label{FInGEN}\\
K_{n}  &  :=\max\left\{  1,2nd_{\text{et}}\frac{1+2\sqrt{\vartheta}%
+2\vartheta}{1-2\sqrt{\vartheta/f_{\mathrm{et}}}}\right\}  ,
\end{align}
where we have set $\vartheta:=(2n)^{-1}\ln(8/\varepsilon)$.

The composable key rate reads~\cite{FSpaper}%
\begin{equation}
R_{\text{gen}}^{\text{het}}\geq\frac{np_{\text{ec}}}{N}\left[
R_{\text{pe,het}}^{(k)}-\frac{\Delta_{\text{aep}}}{\sqrt{n}}+\frac{\Theta
-\Phi_{n}}{n}\right]  , \label{compoHETgen}%
\end{equation}
where $R_{\text{pe,het}}^{(k)}$ is the rate in Eq.~(\ref{RpeFORMU}) depending
on the noise model for the receiver and suitably specified for the heterodyne
protocol. Assuming that the original protocol had $\varepsilon$-security
against collective Gaussian attacks, the symmetrized protocol has security
$\varepsilon^{\prime}=K_{n}^{4}\varepsilon/50$\ against general attacks. Note
that this implies a very demanding condition for the epsilon parameters, such
as $\varepsilon_{\text{pe}}$. As a matter of fact, $\varepsilon_{\text{pe}}$
should be so small that the confidence parameter needs to be calculated
according to Eq.~(\ref{wTAIL}).

\subsection{Numerical investigations}

We may use the previous formulas to plot the composable key rate for the
homodyne/heterodyne protocol with TLO/LLO under each noise model for the
receiver, i.e., corresponding to each of the three different assumptions for
Eve as depicted in Fig.~\ref{attackPIC}. Here we numerically investigate the
most interesting case which is the heterodyne protocol with LLO, for which we
show the performances associated with the three noise models under collective
attacks, and also the worst-case performance associated with the
untrusted-noise model under general attacks. We adopt the physical parameters
listed in Table~\ref{tableMODELS} and the protocol parameters in
Table~\ref{tablePARAMETERS}. The results are given in terms of secret key rate
versus total loss in the channel and can be applied to both fiber-based and
free-space quantum communications, as long as for the latter scenario we can
assume a stable channel (i.e., we can exclude or suitably ignore
fading~\cite{PanosFading}).\begin{table}[t]
\vspace{0.2cm}
\begin{tabular}
[c]{|l|l|l|}\hline
Physical parameter & Symbol & Value\\\hline\hline
Wavelength & $\lambda$ & $800~$nm\\\hline
Detector shot-noise & $\nu_{\text{det}}$ & $2~\text{(het)}$\\\hline
Detector efficiency & $\eta_{\text{eff}}$ & $0.7$ ($1.55~$dB)\\\hline
Detector bandwidth & $W$ & $100~$MHz\\\hline
Noise equivalent power & NEP & $6~$pW/$\sqrt{\text{Hz}}$\\\hline
Linewidth & $l_{\text{W}}$ & $1.6$~KHz\\\hline
LO power & $P_{\text{LO}}$ & $100~$mW\\\hline
Clock & $C$ & $5~$MHz\\\hline
Pulse duration & $\Delta t,\Delta t_{\text{LO}}$ & $10~$ns\\\hline
Setup noise (LLO) & $%
\begin{array}
[c]{l}%
\bar{n}_{\text{ex}}\\
\xi_{\text{ex}}%
\end{array}
$ & $%
\begin{array}
[c]{l}%
\text{Eq.~(\ref{setupNoiseExp})}\\
\text{Eq.~(\ref{excess})}%
\end{array}
$\\\hline
Channel noise & $%
\begin{array}
[c]{l}%
\bar{n}_{B}\\
\xi_{\text{ch}}%
\end{array}
$ & $%
\begin{array}
[c]{l}%
1/500\\
\text{Eq.~(\ref{chexcess})}%
\end{array}
$\\\hline
Total thermal noise & $%
\begin{array}
[c]{l}%
\bar{n}\\
\xi_{\text{tot}}%
\end{array}
$ & $%
\begin{array}
[c]{l}%
\text{Eq.~(\ref{nBARvalue})}\\
\text{Eq.~(\ref{totexcess})}%
\end{array}
$\\\hline
\end{tabular}
\caption{Physical parameters.}%
\label{tableMODELS}%
\end{table}\begin{table}[t]
\vspace{0.2cm}
\begin{tabular}
[c]{|l|l|l|l|}\hline
$%
\begin{array}
[c]{l}%
\text{Protocol}\\
\text{parameter}%
\end{array}
$ & Symbol & $%
\begin{array}
[c]{l}%
\text{Collective}\\
\text{attacks}%
\end{array}
$ & $%
\begin{array}
[c]{l}%
\text{General}\\
\text{attacks}%
\end{array}
$\\\hline\hline
Total pulses & $N$ & $10^{7}$ & $10^{7}$\\\hline
PE signals & $m$ & $0.1\times N$ & $0.1\times N$\\\hline
Energy tests & $f_{\text{et}}$ & $-$ & $0.2$\\\hline
KG signals & $n$ & $0.9\times N$ & $\simeq7.5\times10^{6}$\\\hline
Digitalization & $d$ & $2^{5}$ & $2^{5}$\\\hline
Rec. efficiency & $\beta$ & $0.95$ & $0.95$\\\hline
EC success prob & $p_{\text{ec}}$ & $0.9$ & $0.1$\\\hline
Epsilons & $\varepsilon_{\text{h,s,\ldots}}$ & $2^{-33}\simeq10^{-10}$ &
$10^{-43}$\\\hline
Confidence & $w$ & $\simeq6.34$ & $\simeq14.07$\\\hline
Security & $\varepsilon,\varepsilon^{\prime}$ & $\simeq5.6\times10^{-10}$ &
$\simeq1.4\times10^{-13}$\\\hline
Modulation & $\mu$ & $10$ & $10$\\\hline
\end{tabular}
\caption{Protocol parameters adopted with respect to collective attacks and
general attacks.}%
\label{tablePARAMETERS}%
\end{table}

\begin{figure}[t]
\vspace{0.2cm}
\par
\begin{center}
\includegraphics[width=0.45\textwidth] {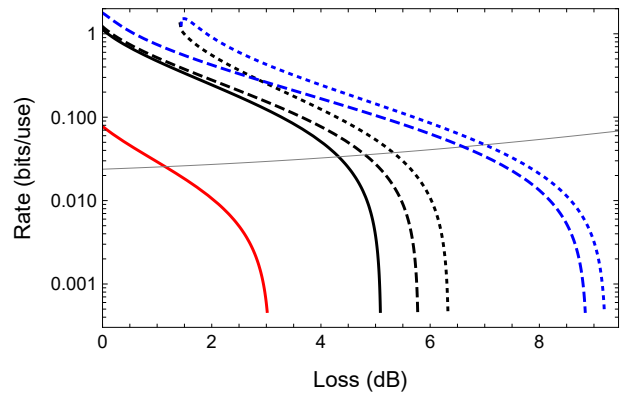}
\end{center}
\par
\vspace{-0.5cm}\caption{Composable secret key rate (bits/use) versus total
loss (decibels) for the heterodyne protocol with LLO. We plot the rates
against collective attacks assuming a trusted-loss and trusted-noise receiver
(black dotted), a trusted-noise receiver (black dashed), and an untrusted
receiver (solid black). We also show the performance achievable with the
untrusted receiver in the presence of general attacks (red). The gray line is
the total excess noise $\xi_{\text{tot}}$ in shot noise units. Finally, the
blue lines refer to line-of-sight security (discussed in Sec.~\ref{LoSsection}%
) for trusted-loss and trusted-noise receiver (blue dotted), and trusted-noise
receiver (blue dashed). Physical and protocol parameters are chosen as in
Tables~\ref{tableMODELS} and~\ref{tablePARAMETERS}.}%
\label{comparisonPIC}%
\end{figure}

The results are shown in Fig.~\ref{comparisonPIC} where we are particularly
interested in the high-rate short-range setting. As we can see from the
figure, the rate has a non-trivial improvement as a result of the stronger
assumptions made for the receiver, as expected. Considering the standard
loss-rate of an optical fiber ($0.2$ dB/km), we see that one extra dB of
tolerance for the rate corresponds to additional $5$ km. Clearly this is
achievable as long as the security assumptions about the receiver are
acceptable by the parties.

\section{Security of near-range free-space quantum communications\label{SEC2}}

Let us now discuss the specific setting of free-space quantum communications
which generally requires some elaborations of the formulas above in order to
account for the additional physical processes occurring in this scenario. In
the following we discuss one potential extra simplification and realistic
assumption for security, and then we treat the issues related to near-range
wireless communications at various frequencies and with different types of
receivers (fixed or mobile).

\subsection{Line-of-sight security\label{LoSsection}}

The line-of-sight (LoS) security is a strong but yet realistic assumption for
free-space quantum communications in the near range (say within $100$ meters
or so). The idea is that transmitter and receiver can \textquotedblleft
see\textquotedblright\ each other, so it is unlikely that Eve is able to
tamper with the middle channel. A realistic attack is here to collect photons
which are lost in the environment; in other words it is a passive attack which
can be interpreted as the action of a pure-loss channel, i.e., a beam-splitter
with no injection of thermal photons (which are the active entangled probes
employed in the usual entangling-cloner attack).

Within the LoS assumption, there are additional degrees of reality for Eve's
attack. The most realistic scenario is Eve using a relatively-small device
which only collects a fraction of the photons that are leaked into the
environment. The worst-case picture which can be used as a bound for the key
is to assume Eve collecting all the leaked photons. In this case, the
performance will strictly depend on how much the receiver is able to intercept
of the incoming beam which is in turn related to the geometric features of the
beam itself (collimated, focused, or spherical beam). In any case, any thermal
noise which is present in the environment is considered to be trusted.

In the studies below, we consider both LoS security (Eve passive on the
channel) and standard security (Eve active on the channel). Under LoS
security, thermal noise is considered to be trusted, which means that the
relevant models for the detector are those with trusted noise [Eve~(2)] and
trusted noise and loss [Eve~(1)]. The attack can be represented as in
Fig.~\ref{attackPIC}\ but where Eve does not control environmental modes,
represented by mode $e$ for Eve~(1) and modes $e$,$v$ for Eve~(2). With the
trusted-noise detector, we also allow Eve to collect leakage from Bob's setup;
with the trusted-noise-and-loss detector, this additional side-channel is
excluded. Depending on the cases, we adopt one assumption or the other. See
Table~\ref{tableSEClevels} for a summary of the security types and trust
levels (associated detector models). These definitions are meant to be in
addition to the classification into individual, collective and
coherent/general attacks.

\begin{table}[h]%
\begin{tabular}
[c]{|l|l|l|}\hline
Channel noise & Security type & Detector model\\\hline\hline
Untrusted &
\begin{tabular}
[c]{l}%
Standard security\\
(Active Eve\\
controlling the\\
environment)
\end{tabular}
&
\begin{tabular}
[c]{l}%
\begin{tabular}
[c]{l}%
$\bullet~$Untrusted\\
\lbrack Eve~(3)]
\end{tabular}
\\%
\begin{tabular}
[c]{l}%
$\bullet$~Noise-trusted\\
\lbrack Eve~(2)]
\end{tabular}
\\%
\begin{tabular}
[c]{l}%
$\bullet~$Noise-loss-trusted\\
\lbrack Eve~(1)]
\end{tabular}
\end{tabular}
\\\hline
Trusted &
\begin{tabular}
[c]{l}%
LoS security\\
(Passive Eve.\\
No control of\\
the environment)
\end{tabular}
&
\begin{tabular}
[c]{l}%
\begin{tabular}
[c]{l}%
$\bullet~$Noise-trusted\\
\lbrack Eve~(2)]
\end{tabular}
\\%
\begin{tabular}
[c]{l}%
$\bullet~$Noise-loss-trusted\\
\lbrack Eve~(1)]
\end{tabular}
\end{tabular}
\\\hline
\end{tabular}
\caption{Security types and trust levels (detector models). The security
assumptions become stronger from top to bottom. }%
\label{tableSEClevels}%
\end{table}

The secret key rates under LoS security are derived by excluding Eve from the
control of the environmental noise. This means that her CM\ is reduced from
the form in Eq.~(\ref{CMcase1}) to just the block $\phi\mathbf{I}$. Thus, we
have to consider the simpler joint CM for Bob and Eve%
\begin{equation}
\mathbf{V}_{BE}=\left(
\begin{array}
[c]{cc}%
b\mathbf{I} & \theta\mathbf{I}\\
\theta\mathbf{I} & \phi\mathbf{I}%
\end{array}
\right)  , \label{VbeLOS}%
\end{equation}
leading to the conditional CMs%
\begin{equation}
\mathbf{V}_{E|B}^{\text{hom}}=\left(
\begin{array}
[c]{cc}%
\phi-\frac{\theta^{2}}{b} & 0\\
0 & \phi
\end{array}
\right)  ,~\mathbf{V}_{E|B}^{\text{het}}=\left(  \phi-\frac{\theta^{2}}%
{b+1}\right)  \mathbf{I}. \label{Vbe2LOS}%
\end{equation}
Therefore, Eve's Holevo bound to be used in the key rates is simply given by%
\begin{align}
\chi_{\text{LoS}}^{\text{hom}}(E  &  :y)=H(\phi)-H\left[  \sqrt{\phi
(\phi-\theta^{2}/b)}\right]  ,\\
\chi_{\text{LoS}}^{\text{het}}(E  &  :y)=H(\phi)-H\left(  \phi-\frac
{\theta^{2}}{b+1}\right)  , \label{HetLOSformula}%
\end{align}
where the explicit expressions for $\theta$ and $\phi$ depend on the detector
noise model, while $b$ is given in Eq.~(\ref{parameterB}).

Using these expressions, we may then write the asymptotic key rate with LoS
security for the two detector models ($k=1,2$). Recalling that the mutual
information is expressed as in Eq.~(\ref{MutualINFOeq}), the LoS key rate is
given by%
\begin{equation}
R_{\text{asy,LoS}}^{(k)}(\tau,\bar{n},\bar{n}_{B})=\beta I(x:y)_{\tau,\bar{n}%
}-\chi_{\text{LoS}}(E:y)_{\tau,\bar{n},\bar{n}_{B}},
\end{equation}
taking specific expressions for the homodyne protocol [$R_{\text{asy,LoS,hom}%
}^{(k)}$] and the heterodyne protocol [$R_{\text{asy,LoS,het}}^{(k)}$]. After
parameter estimation, the modified key rate will be expressed in terms of the
worst-case estimators as $R_{\text{pe,LoS}}^{(k)}=R_{\text{asy,LoS}}%
^{(k)}(\tau^{\prime},\bar{n}^{\prime},\bar{n}_{B}^{\prime})$. Finally, the
composable finite-size LoS key rate takes the expression in Eq.~(\ref{sckeee})
proviso we make the replacement $R_{\text{pe}}^{(k)}\longrightarrow
R_{\text{pe,LoS}}^{(k)}$. Improvement in performance is shown in
Fig.~\ref{comparisonPIC}.

\subsection{Optical wireless with fixed devices}

Let us consider a free-space optical link between transmitter and receiver.
Assume that this is mediated by a Gaussian \textrm{TEM}$_{00}$ beam with
initial spot-size $w_{0}$ and phase-front radius of curvature $R_{0}%
$~\cite{Goodman,Siegman,svelto}. This beam has a single well-defined
polarization (scalar approximation) and carrier frequency $\nu=c/\lambda$,
with $\lambda$ being the wavelength and $c$ the speed of light (so angular
frequency is $\omega=2\pi c/\lambda$, and wavenumber is $k=\omega
/c=2\pi/\lambda$). The pulse duration $\Delta t$ and frequency bandwidth
$\Delta\nu$ satisfy the time-bandwidth product for Gaussian pulses, i.e.,
$\Delta t\Delta\nu\gtrsim0.44$. In particular, we may assume $\Delta
t\Delta\nu\simeq1$. Under the paraxial wave approximation, we assume
free-space propagation along the $z$ direction with no limiting apertures in
the transverse plane, neglecting diffraction effects at the transmitter (e.g.,
by assuming a suitable aperture for the transmitter with radius $\geq2w_{0}%
$~\cite{Siegman}).

By introducing the Rayleigh range
\begin{equation}
z_{R}:=\frac{\pi w_{0}^{2}}{\lambda},
\end{equation}
which identifies near- and far-field, we may write the following expression
for the diffraction-limited spot size of the beam at generic distance
$z$~\cite{Siegman,svelto}%
\begin{equation}
w_{z}^{2}=w_{0}^{2}\left[  \left(  1-\frac{z}{R_{0}}\right)  ^{2}+\left(
\frac{z}{z_{R}}\right)  ^{2}\right]  .
\end{equation}
In particular, for a collimated beam ($R_{0}=\infty$), we get
\begin{equation}
w_{z}^{2}=w_{0}^{2}[1+(z/z_{R})^{2}], \label{colliB}%
\end{equation}
while for a focused beam ($R_{0}=z$), we have%
\begin{equation}
w_{z}^{2}=w_{0}^{2}(z/z_{R})^{2}=\left(  \frac{\lambda z}{\pi w_{0}}\right)
^{2}. \label{focusedB}%
\end{equation}
We see that, in the far field $z\gg z_{R}$, the expressions in
Eqs.~(\ref{colliB}) and~(\ref{focusedB}) tend to coincide.

Consider then a receiver with a sharped-edged circular aperture with radius
$a_{R}$. The total power impinging on this aperture is given by%
\begin{equation}
P(z,a_{R})=\frac{\pi w_{0}^{2}}{2}\eta_{\text{d}},~\eta_{\text{d}%
}:=1-e^{-2a_{R}^{2}/w_{z}^{2}}, \label{ILexp}%
\end{equation}
where parameter $\eta_{\text{d}}$ is the non-unit transmissivity of the
channel due to the free-space diffraction and the finite size of the receiver.
Note that, for far field and a receiver's size comparable with the
transmitter's (so $a_{R}\simeq w_{0}$), we have $w_{z}\gg a_{R}$ and therefore
the approximation
\begin{equation}
\eta_{\text{d}}\simeq\eta_{\text{d}}^{\text{far}}:=2a_{R}^{2}/w_{z}^{2}\ll1.
\end{equation}
For a collimated or focused beam, this becomes%
\begin{equation}
\eta_{\text{d}}^{\text{far}}\simeq2\left(  \frac{\pi w_{0}a_{R}}{\lambda
z}\right)  ^{2}.
\end{equation}

The overall transmissivity of the system can be written as $\tau
=\eta_{\text{ch}}\eta_{\text{eff}}$, where $\eta_{\text{ch}}=\eta_{\text{d}%
}\eta_{\text{atm}}$ is the total transmissivity of the external channel which
generally includes the effect of atmospheric extinction $\eta_{\text{atm}}$.
Since the latter effect is negligible at short distances ($\eta_{\text{atm}%
}\simeq1$), we may just write $\eta_{\text{ch}}\simeq\eta_{\text{d}}$. By
contrast, the other term $\eta_{\text{eff}}$ is the total quantum efficiency
of the receiver and its contribution is typically non-negligible, e.g.,
$\eta_{\text{eff}}\simeq0.7$. Because the devices are assumed to be fixed,
there is no fading, meaning that the total transmissivity can be assumed to be
constant and equal to $\tau$.

The quantum communication scenario can be described as in Fig.~\ref{attackPIC}%
, where $\eta_{\text{ch}}$ is essentially given by free-space diffraction and
the thermal background $\bar{n}_{B}$ needs to be carefully evaluated from the
sky brightness (see below). Then, we can certainly assume standard security
with the trust levels $k=0,1,2$ according to which Eve's interaction is
described by different effective beam-splitters with different amounts of
input thermal noise $\bar{n}_{e}^{(k)}$ (see Sec.~\ref{Section_levels}).
Similarly, we may investigate LoS security where thermal noise is assumed to
be trusted.

Sky brightness $B_{\lambda}^{\text{sky}}$ is measured in W m$^{-2}$ nm$^{-1}$
sr$^{-1}$ and its value typically varies from $\simeq1.5\times10^{-6}$ (clear
night)\ to $\simeq1.5\times10^{-1}$ (cloudy day)~\cite{Miao,BrussSAT}, if one
assumes that the receiver field of view is shielded from direct exposition to
bright sources (e.g., the sun). Let us assume a receiver with aperture $a_{R}$
and angular field of view $\Omega_{\text{fov}}$ (in steradians). Assume the
receiver has a detector with bandwidth $W$ and spectral filter $\Delta\lambda
$. Then, the mean number of background thermal photons per mode collected by
the receiver is equal to%
\begin{equation}
\bar{n}_{B}=\frac{\pi\lambda\Gamma_{R}}{hc}B_{\lambda}^{\text{sky}}%
,~\Gamma_{R}:=\Delta\lambda W^{-1}\Omega_{\text{fov}}a_{R}^{2}. \label{downT}%
\end{equation}
In this formula, we can estimate $\Omega_{\text{fov}}^{1/2}\simeq
2\arctan[l_{\text{D}}/(2f_{\text{D}})]$ from the linear size of the sensor of
the detector $l_{\text{D}}$ and the focal length $f_{\text{D}}$ of the
receiver. For $l_{\text{D}}=2~$mm and $f_{\text{D}}=20~$cm, we find
$\Omega_{\text{fov}}\simeq10^{-4}~$sr. Note that the latter value of the field
of view is relatively-large compared with typical values considered in
long-range setting, including satellite communications (where $\Omega
_{\text{fov}}\simeq10^{-10}~$sr).

The effective value of the spectral filter $\Delta\lambda$ can be very narrow
in setups that are based on homodyne/heterodyne detection. The reason is
because the required mode-matching of the signal with the LO pulse provides a
natural interferometric process which effectively reduces the filter
potentially down to the time-product bandwidth. For instance, for an LO\ pulse
of $\Delta t_{\text{LO}}=10~$ns, we may assume a bandwidth $\Delta\nu=50~$MHz
which is $\geq0.44/\Delta t_{\text{LO}}$. Thus, interferometry at the homodyne
setup imposes an effective filter of $\Delta\lambda=\lambda^{2}\Delta
\nu/c\simeq0.1~$pm around $\lambda=800~$nm.

Finally, if we take the detector bandwidth $W=100~$MHz and we assume a small
area for the receiver's aperture, i.e., $a_{R}=1~$cm (so as to be compatible
with the typical sizes of near-range devices), then we compute $\bar{n}%
_{B}\simeq0.019$ photons per mode during a cloudy day. This is a non-trivial
amount of noise that leads to a clear discrepancy between the performance in
standard security (where channel's noise is considered to be untrusted) and
LoS security (where this noise is assumed to be trusted). Let us also remark
here that LoS security is a realistic assumption for receivers with a small
field of view, so the noise collected from free space is limited and unlikely
to come from an active Eve hidden in the environment.

For our numerical study we consider the physical parameters listed in
Table~\ref{WirelessTABLE}; these are compatible with indoor and near-range
optical wireless communications with small devices (e.g., laptops). This means
that, for the transmitter, we consider limited power (e.g., $10~$mW), and a
small spot size ($w_{0}=1~$mm). Similarly, for the receiver, we consider a
limited aperture ($a_{R}=1~$cm), non-unit quantum efficiency ($\eta
_{\text{eff}}=0.7$), and a realistic field of view $\Omega_{\text{fov}}%
\simeq10^{-4}~$sr as discussed above.

\begin{figure}[t]
\vspace{0.1cm}
\par
\begin{center}
\includegraphics[width=0.45\textwidth] {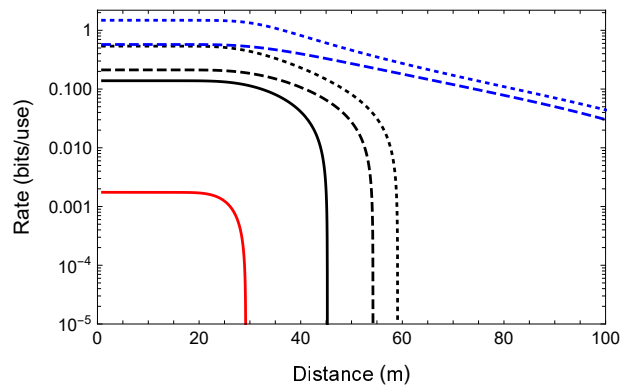}
\end{center}
\par
\vspace{-0.5cm}\caption{Optical-wireless QKD\ with fixed devices. We plot the
composable secret key rate (bits/use) versus free-space distance (meters) for
the heterodyne protocol with LLO. In particular, we show the rates against
collective attacks assuming a trusted-loss-and-noise receiver (black dotted),
a trusted-noise receiver (black dashed), and an untrusted receiver (solid
black). We also show the performance achievable with the untrusted receiver
versus general attacks (red). The blue lines refer to line-of-sight security
(discussed in Sec.~\ref{LoSsection}) for trusted-loss-and-noise receiver (blue
dotted), and trusted-noise receiver (blue dashed). Physical parameters are
chosen as in Table~\ref{WirelessTABLE}, while protocol parameters are in
Table~\ref{tablePARAMETERS}.}%
\label{FixedWirelessPic}%
\end{figure}

\begin{table}[t]%
\begin{tabular}
[c]{|l|l|l|}\hline
Physical parameter & Symbol & Value\\\hline\hline
Altitude & $h$ & $30$~m\\\hline
Beam curvature & $R_{0}$ & $\infty$ (collimated)\\\hline
Wavelength & $\lambda$ & $800~$nm\\\hline
Beam spot size & $w_{0}$ & $1~$mm\\\hline
Receiver aperture & $a_{R}$ & $1$ cm\\\hline
Receiver field of view & $\Omega_{\text{fov}}$ & $10^{-4}~$sr\\\hline
Homodyne filter & $\Delta\lambda$ & $0.1~\text{pm}$\\\hline
Detector shot-noise & $\nu_{\text{det}}$ & $2~\text{(het)}$\\\hline
Detector efficiency & $\eta_{\text{eff}}$ & $0.7$ ($1.55~$dB)\\\hline
Detector bandwidth & $W$ & $100~$MHz\\\hline
Noise equivalent power & NEP & $6~$pW/$\sqrt{\text{Hz}}$\\\hline
Linewidth & $l_{\text{W}}$ & $1.6$~KHz\\\hline
LO power & $P_{\text{LO}}$ & $10~$mW\\\hline
Clock & $C$ & $5~$MHz\\\hline
Pulse duration & $\Delta t,\Delta t_{\text{LO}}$ & $10~$ns\\\hline
Setup noise with LLO & $\bar{n}_{\text{ex}}$ & Eq.~(\ref{setupNoiseExp}%
)\\\hline
Channel noise & $\bar{n}_{B}$ & $0.019$ [Eq.~(\ref{downT})]\\\hline
Total thermal noise & $\bar{n}$ & Eq.~(\ref{nBARvalue})\\\hline
Atmospheric extinction & $\eta_{\text{atm}}$ & $\simeq1$ (negligible)\\\hline
\end{tabular}
\caption{Physical parameters for optical wireless}%
\label{WirelessTABLE}%
\end{table}Assuming the physical parameters in Table~\ref{WirelessTABLE} and
the protocols parameters in Table~\ref{tablePARAMETERS}, we show the various
achievable performances of the free-space diffraction-limited heterodyne
protocol with LLO in Fig.~\ref{FixedWirelessPic}. As we can see from the
figure, we have drastically different rates depending on the type of security
and trust level. It is clear that the highest rates (and distances) are
obtained with LoS security (blue lines in the figure). With standard security,
the range is restricted to about $50$ meters (black lines in the figure) and
about $30$ meters in the worst-case scenario of an untrusted detector and
general attacks (red line in the figure). The possibility to enforce weaker
security assumptions leads to non-trivial advantages in terms of rate and
distance.

Also note the stability of the rates at short distances ($<30~$m) where their
values remain approximately constant. This is due to the fact that, for the
specific regime of parameters considered, the beam broadening induced by
free-space diffraction within that range [see Eq.~(\ref{colliB}) with
$w_{0}=1~$mm and $z<30$~m] is still limited with respect to the radius of the
receiver's aperture ($a_{R}=1~$cm). Thus, the transmissivity $\eta_{\text{d}}$
in Eq.~(\ref{ILexp}) remains sufficiently close to $1$, before starting to
decay after about $30$~m.

\subsection{Optical wireless with mobile devices}

\subsubsection{Pointing and tracking error}

In the presence of free-space optical connections with portable devices, one
can use a suitable tracking mechanism so the transmitter (such as a fixed
router/hot spot) points at the mobile receiver in real time with some small
pointing error. In general, the receiver too may have a mechanism of adaptive
optics aimed at maintaining the beam alignment by rotating the field of view
in direction of the transmitter. We therefore need to introduce a pointing
error at the transmitter $\tilde{\sigma}_{\text{P}}$ which introduces a
Gaussian wandering of the beam centroid over the receiver's aperture with
variance $\sigma_{\text{P}}^{2}\simeq(\tilde{\sigma}_{\text{P}}z)^{2}$ for
distance $z$. We assume an accessible value $\tilde{\sigma}_{\text{P}}%
\simeq1.745\times10^{-3}$ radiant, which is about $1/10$ of a degree (this is
orders-of-magnitude worse than the performance achievable in satellite-based
pointing and tracking).

Let us call $r$ the instantaneous deflection of the beam centroid from the
center of the receiver's aperture. The wandering can be described by the
Weibull distribution
\begin{equation}
P_{\text{WB}}(r)=\frac{r}{\sigma_{\text{P}}^{2}}\exp\left(  -\frac{r^{2}%
}{2\sigma_{\text{P}}^{2}}\right)  . \label{WeibullDISTRIBUTION}%
\end{equation}
For each value of the deflection $r$, there is an associated instantaneous
transmissivity $\tau=\tau(r)$, which can be computed as follows%
\begin{equation}
\tau(r)=e^{-\frac{4r^{2}}{w_{z}^{2}}}Q_{0}\left(  \frac{2r^{2}}{w_{z}^{2}%
},\frac{4ra_{R}}{w_{z}^{2}}\right)  , \label{etadiff}%
\end{equation}
where $Q_{0}(x,y)$ is an incomplete Weber integral~\cite{Agrest}.

Alternatively, we may use the approximation
\begin{equation}
\tau(r)\simeq\eta\exp\left[  -\left(  \frac{r}{r_{0}}\right)  ^{\gamma
}\right]  , \label{etaSTanalytical}%
\end{equation}
where%
\begin{equation}
\eta:=\tau(0)=\eta_{\text{ch}}(z)\eta_{\text{eff}}\simeq\eta_{\text{d}}%
(z)\eta_{\text{eff}} \label{eta0}%
\end{equation}
is the maximum transmissivity at distance $z$ (corresponding to a beam that is
perfectly-aligned), while$\ \gamma$ and $r_{0}$\ are the following shape and
scale (positive) parameters
\begin{align}
\gamma &  =\frac{4\eta_{\text{d}}^{\text{far}}\Lambda_{1}(\eta_{\text{d}%
}^{\text{far}})}{1-\Lambda_{0}(\eta_{\text{d}}^{\text{far}})}\left[  \ln
\frac{2\eta_{\text{d}}}{1-\Lambda_{0}(\eta_{\text{d}}^{\text{far}})}\right]
^{-1},\label{expGG1}\\
r_{0}  &  =a_{R}\left[  \ln\frac{2\eta_{\text{d}}}{1-\Lambda_{0}%
(\eta_{\text{d}}^{\text{far}})}\right]  ^{-\frac{1}{\gamma}}, \label{expGG2}%
\end{align}
where $\Lambda_{n}(x):=e^{-2x}I_{n}\left(  2x\right)  $ and $I_{n}$ is a
modified Bessel function of the first kind with order $n$~\cite[Eq.~(D2)]%
{Vasy12}.

By suitably combining Eqs.~(\ref{WeibullDISTRIBUTION})
and~(\ref{etaSTanalytical}), one can derive the fading statistics, i.e., the
probability distribution $P_{\text{fad}}$ associated with the instantaneous
transmissivity $\tau$, which is given by%

\begin{equation}
P_{\text{fad}}(\tau)=\frac{r_{0}^{2}}{\gamma\sigma_{\text{P}}^{2}\tau}\left(
\ln\frac{\eta}{\tau}\right)  ^{\frac{2}{\gamma}-1}\exp\left[  -\frac{r_{0}%
^{2}}{2\sigma_{\text{P}}^{2}}\left(  \ln\frac{\eta}{\tau}\right)  ^{\frac
{2}{\gamma}}\right]  . \label{P0tau}%
\end{equation}

\subsubsection{Maximum wireless range}

Besides the beam wandering (and associated fading) due to pointing and
tracking error, there is also the further issue that a mobile receiver
generally has a variable distance from the transmitter, so the transmissivity
of the free-space link has an additional degree of variability. The latter
effect has a very slow dynamics with respect to typical clocks, meaning that a
block of reasonable size is distributed while the position of the receiver is
substantially unchanged. For example, for a detector bandwidth $W=100~$MHz, we
may use a clock of $C=W/3\simeq33~$MHz. In this case, a block of $10^{7}$
points will be distributed in $1/3$ of a second. For an indoor network,
assuming an average walking speed of $\simeq1.5$ m/s, this corresponds to a
$\simeq50$~cm free-space displacement of the receiver. In the worst-case
scenario where this displacement increases the distance from the transmitter,
we may assume that the distribution of the whole block occurs at the maximum distance.

In general, we may compute a lower bound by assuming that the entire quantum
communication (i.e., the communication of all the blocks) occurs with the
mobile device at the maximum distance from the transmitter. In other words, we
can fix a maximum range $z_{\text{max}}$ for the local network and assume this
value as worst-case scenario. Since the parties control the parameters of the
channel and know the instantaneous distance, they could process their data in
a way that it appears to be completely distributed at $z_{\text{max}}$ (data
distributed at $z<z_{\text{max}}$ can be attenuated and suitably thermalized
in post processing).

To be more precise the lower bound should be computed by minimizing the
transmissivity and maximizing the thermal noise over the distance $z\leq
z_{\text{max}}$, so that data is processed via a more lossy and noisy
channel. While the minimization of the transmissivity occurs at
$z=z_{\text{max}}$, the maximization of the thermal noise may occur at
different values of $z$, depending on the type of LO. In particular, this
value is $z=z_{\text{max}}$ for the TLO and $z=0$ for the LLO. The issue is
therefore resolved for the LLO if we keep the mobile device at
$z=z_{\text{max}}$ while bounding the LLO noise with the value for $z=0$.

Such an approach is not optimal but robust and applicable to outdoor wireless
networks with faster-moving devices (with a speed limited by the ratio between
$z_{\text{max}}$ and the total communication time). It is worth mentioning
that, a better but more complicated strategy relies on slicing the trajectory
of the moving device into sectors, with each sector being associated with the
communication of a single block and the final rate being given by the average
rate over the sectors. This is particularly useful in satellite quantum
communications where a trajectory is well defined (for instance, see the
technique of orbital slicing in Ref.~\cite{SATpaper}). However, for stochastic
trajectories on the ground, the analytical treatment is not immediate.

\subsubsection{Pilot modes and de-fading\label{defadingSEC}}

Besides the use of bright pointing/tracking modes and bright LLO-reference
modes, it is also important to use relatively-bright pilot modes that are
specifically employed for the real-time estimation of the instantaneous
transmissivity $\tau$, whose fluctuation is generally due to both pointing
error and distance variability (for mobile devices). These $m_{\text{PL}}$
pilots are randomly interleaved with $N_{\text{S}}:=N-m_{\text{PL}}$ signal
modes, where $N$\ are the total pulses. The pilots allow the parties to:
(i)~identify an overall interval for the transmissivity $\Delta=[\tau_{\min
},\tau_{\max}]$ in which $N_{\text{S}}p_{\Delta}$ signals are post-selected
with probability $p_{\Delta}$; (ii)~introduce a lattice in $\Delta$ with step
$\delta\tau$, so that each signal is associated with a corresponding narrow
bin of transmissivities $\Delta_{k}:=[\tau_{k},\tau_{k+1}]$, with $\tau
_{k}:=\tau_{\min}+(k-1)\delta\tau$ for $k=1,\ldots,M$ and $M=(\tau_{\max}%
-\tau_{\min})/\delta\tau$~\cite{NotePilots}.

Each bin $\Delta_{k}$ is selected with probability $p_{k}$\ and, therefore,
populated by $N_{\text{S}}p_{k}$ signals. There are corresponding
$\nu_{\text{det}}N_{\text{S}}p_{k}$ pairs of points $\{x_{i},y_{i}%
\}$\ satisfying the input-output relation of Eq.~(\ref{IOnew}), which here
reads%
\begin{equation}
y^{(k)}\simeq\sqrt{\tau_{k}}x+z^{(k)},
\end{equation}
where $z^{(k)}$ is a Gaussian noise variable with variance%
\begin{equation}
\sigma_{k}^{2}=2\bar{n}_{k}+\nu_{\text{det}},~\bar{n}_{k}:=\eta_{\text{eff}%
}\bar{n}_{B}+\bar{n}_{\text{ex}}(\tau_{k}).
\end{equation}
Bob can map these points into the first bin $\Delta_{1}$ of the interval via
the de-fading map%
\begin{equation}
y^{(k)}\rightarrow\tilde{y}^{(k)}=\sqrt{\frac{\tau_{\text{min}}}{\tau_{k}}%
}y^{(k)}+\sqrt{1-\frac{\tau_{\text{min}}}{\tau_{k}}}\xi_{\text{add}},
\end{equation}
where $\xi_{\text{add}}$ is Gaussian noise with variance $\nu_{\text{det}}$.

By repeating this procedure for all the bins, Bob create the new variable
\begin{equation}
\tilde{y}=\sqrt{\tau_{\text{min}}}x+\tilde{z}, \label{finalIO}%
\end{equation}
where $\tilde{z}$ is non-Gaussian noise with variance%
\begin{equation}
\sigma_{\tilde{z}}^{2}=2\bar{n}_{\ast}+\nu_{\text{det}},~\bar{n}_{\ast}%
:=\frac{\tau_{\text{min}}}{p_{\Delta}}\sum_{k}\frac{p_{k}}{\tau_{k}}\bar
{n}_{k}. \label{thermalNOISE}%
\end{equation}
This new variable is now associated with a single (worst-case) transmissivity
$\tau_{\text{min}}$, thus effectively removing the fading process from the
distributed data, i.e., from their $\nu_{\text{det}}N_{\text{S}}p_{\Delta}$
pairs of correlated points.

Exploiting the optimality of Gaussian attacks, the parties assume that
$\tilde{z}$ is Gaussian (overestimating Eve's performance). In this way, the
final input-output relation in Eq.~(\ref{finalIO}) reduces to considering a
simpler thermal-loss\ Gaussian channel with transmissivity $\tau_{\text{min}}$
and thermal number $\bar{n}_{\ast}$. See Ref.~\cite{FSpaper} for more details.

For a receiver at some fixed distance $z$ and only subject to pointing error,
we can assume $\tau_{\max}=\eta$ [cf. Eq.~(\ref{eta0})] and $\tau_{\text{min}%
}=f_{\text{th}}\eta$ for some threshold factor $f_{\text{th}}<1$. Then, the
probabilities $p_{\Delta}=p(\tau_{\text{min}},\tau_{\text{max}})$ and
$p_{k}=p(\tau_{k},\tau_{k+1})$ are computed from the formula%
\begin{equation}
p(\tau_{1},\tau_{2}):=\int_{\tau_{1}}^{\tau_{2}}d\tau~P_{\text{fad}}(\tau),
\label{probFORMULA}%
\end{equation}
where $P_{\text{fad}}(\tau)$ is given in Eq.~(\ref{P0tau}).

In general, for a mobile receiver at variable distance $z$, Alice and Bob
compute the post-selection interval $\Delta$ and the lattice $\{\Delta_{k}%
\}$\ directly from data, together with the corresponding values of $p_{\Delta
}$ and $p_{k}$. As mentioned in the previous subsection, the performance in
this general scenario can be lower-bounded by the extreme case where the
receiver is assumed to be fixed at the maximum distance $z_{\text{max}}$ from
the transmitter (while maximizing thermal noise over $z$, whose maximum is at
$z_{\text{max}}$ for a TLO and at $z=0$ for an LLO). In this worst-case
scenario, we may exploit the formula in Eq.~(\ref{probFORMULA}) for the fading
probability (suitably computed at $z_{\text{max}}$) and derive an analytical
lower bound for the secret key rate.

\subsubsection{Estimators and key rate}

Let us assume the worst-case scenario of a receiver at the maximum range
$z_{\text{max}}$ from the transmitter, so the maximum transmissivity is
$\tau_{\max}=\eta(z_{\text{max}})$ and the minimum transmissivity is
$\tau_{\text{min}}=f_{\text{th}}\eta(z_{\text{max}})$ for some threshold value
$f_{\text{th}}$. These border values define a post-selection interval $\Delta$
which is sliced into a lattice of $M$\ narrow bins $\{\Delta_{k}\}$. The
instantaneous transmissivity $\tau$ will fluctuate according to the
distribution in Eq.~(\ref{P0tau}) with associated pointing error
$\sigma_{z_{\text{max}}}^{2}\simeq(\sigma_{\text{P}}z_{\text{max}})^{2}$ for
an empirical value $\sigma_{\text{P}}$ at the transmitter (e.g., $1/10$ of a
degree). As a result of the fluctuation, a value of the transmissivity $\tau$
is post-selected with probability $p_{\Delta}$ and populates bin $\Delta_{k}$
with probability $p_{k}$, according to the integral in Eq.~(\ref{probFORMULA}).

For the worst-case scenario, let us also assume that the thermal noise is
maximized over $z\leq z_{\text{max}}$ (and the fading process). Thus, for any
bin $\Delta_{k}$, we consider the following bound on the associated thermal
noise%
\begin{equation}
\bar{n}_{k}\leq\bar{n}_{\text{wc}}=\eta_{\text{eff}}\bar{n}_{B}+\bar
{n}_{\text{ex,wc}}, \label{nboundd}%
\end{equation}
where the maximum setup noise $\bar{n}_{\text{ex,wc}}$ depends on the type of
LO and is given by%
\begin{equation}
\bar{n}_{\text{ex,wc}}^{\text{TLO}}\simeq\Theta_{\text{el}}/\tau_{\text{min}%
},~\bar{n}_{\text{ex,wc}}^{\text{LLO}}\simeq\Theta_{\text{el}}+\pi\sigma
_{x}^{2}C^{-1}l_{\text{W}}. \label{eqabove}%
\end{equation}
Note that the first expression in Eq.~(\ref{eqabove}) above is computed on
$\tau_{\text{min}}=\tau_{\text{min}}(z_{\text{max}})$ while the second one is
computed for $\tau=1$ (maximum value at $z=0$). By replacing
Eq.~(\ref{nboundd}) in Eq.~(\ref{thermalNOISE}), we get the bound
\begin{equation}
\bar{n}_{\ast}\leq\bar{n}_{\text{wc}}.
\end{equation}

As already explained, the construction of the lattice is possible thanks to
the random pilots. In total, during the quantum communication, the parties
exchange\ $N$ quantum pulses, whose $m_{\text{PL}}$ are pilots and
$N_{\text{S}}=N-m_{\text{PL}}$ are signals. Using the pilots, the parties
post-select a fraction $N_{\text{S}}p_{\Delta}$ of the signals, with a smaller
fraction $N_{\text{S}}p_{k}$ allocated to the generic bin $\Delta_{k}$. After
de-fading, the parties are connected by an effective thermal-loss channel with
transmissivity $\tau_{\text{min}}=\tau_{\text{min}}(z_{\text{max}})$ and
thermal number $\bar{n}_{\text{wc}}$.

The parties sacrifice a portion $mp_{\Delta}$ of the post-selected signals
$N_{\text{S}}p_{\Delta}$ for parameter estimation (PE), so $np_{\Delta}$
signals are left for key generation, where $n=N_{\text{S}}-m$ (this value is
further reduced for security extended to general coherent attacks). Overall
the parties use $m_{\Delta}:=\nu_{\text{det}}mp_{\Delta}$ pairs of data points
for PE following the procedure described in Sec.~\ref{PEsection} with
effective transmissivity $\tau_{\text{min}}=\tau_{\text{min}}(z_{\text{max}})$
and $\sigma_{\text{wc}}^{2}=2\bar{n}_{\text{wc}}+\nu_{\text{det}}$. This leads
to the following bounds for the worst-case estimators~\cite{FSpaper}%
\begin{align}
\tau_{\text{LB}}  &  =\tau_{\text{min}}-2w\sqrt{\frac{2\tau_{\text{min}}%
^{2}+\tau_{\text{min}}\sigma_{\text{wc}}^{2}/\sigma_{x}^{2}}{m_{\Delta}}},\\
\bar{n}_{\text{UB}}  &  =\bar{n}_{\text{wc}}+w\frac{\sigma_{\text{wc}}^{2}%
}{\sqrt{2m_{\Delta}}},
\end{align}
where $\sigma_{x}^{2}$ is the input modulation and $w$ is the confidence
parameter [cf. Eqs.~(\ref{wSTVALUE}) and~(\ref{wTAIL})].

As we can see from the two estimators above, the relevant information is the
minimum transmissivity $\tau_{\text{min}}$ of the post-selection interval, the
maximum thermal noise $\bar{n}_{\text{wc}}$ over the range (and fading
process), and the number of post-selected points $m_{\Delta}$. The formulas
hold for a generic fading statistics, i.e., not necessarily given by
Eq.~(\ref{probFORMULA}), as long as we can evaluate $m_{\Delta}$. Also note
that, assuming Eq.~(\ref{probFORMULA}) and fixing a threshold transmissivity
$\tau_{\text{min}}$, the value of $m_{\Delta}$\ decreases by increasing $z$.
In other words, the fact that a worst-case device at the maximum range
provides a lower bound for a mobile device is also due to the decreased
statistics for PE.

%%%%%%%%%%%%%%%%%%%%%%%%%%%%%%%%%%%%%%%%%%%%%%%%%%%%%%%%%%%%%%%%%%%%%%%%%%%%%%%%%%%%%%%%%%%%%%%%%%%% formulas, table with parameters, plots

In order to compute the key rates for the trusted models, we also need to
bound the worst-case estimator of the background thermal noise $\bar{n}_{B}$.
This is possible by writing%
\begin{equation}
\bar{n}_{B}^{\text{UB}}=\frac{\bar{n}_{\text{UB}}-\bar{n}_{\text{ex,bc}}}%
{\eta_{\text{eff}}},
\end{equation}
where the best-case value $\bar{n}_{\text{ex,bc}}$ needs to be optimized over
the entire range $z\leq z_{\max}$ and the fading process. We therefore extend
Eqs.~(\ref{bcc1}) and~(\ref{bcc2}) to the following expressions%
\begin{equation}
\bar{n}_{\text{ex,bc}}^{\text{TLO}}:=\Theta_{\text{el}},~\bar{n}%
_{\text{ex,bc}}^{\text{LLO}}:=\Theta_{\text{el}}+\Theta_{\text{ph}}%
\tau_{\text{min}}.
\end{equation}

We now have all the elements to write the composable finite-size key rate,
which extends Eq.~(\ref{sckeee}) of Sec.~\ref{KR_sec} to the following
expression%
\begin{equation}
R\geq\frac{np_{\Delta}p_{\text{ec}}}{N}\left(  R_{\text{pe}}^{(k)}%
-\frac{\Delta_{\text{aep}}}{\sqrt{np_{\Delta}}}+\frac{\Theta}{np_{\Delta}%
}\right)  , \label{erreDELTAA}%
\end{equation}
where $n=N-(m+m_{\text{PL}})$ and $R_{\text{pe}}^{(k)}$ depends on the
receiver model ($k=1,2,3$). The latter takes the following expressions in
terms of the new estimators%
\begin{align}
R_{\text{pe}}^{(1,2)}  &  =R_{\text{asy}}^{(1,2)}(\tau_{\text{LB}},\bar
{n}_{\text{UB}},\bar{n}_{B}^{\text{UB}}),\label{RpeFREE1}\\
R_{\text{pe}}^{(3)}  &  =R_{\text{asy}}^{(3)}(\tau_{\text{LB}},\bar
{n}_{\text{UB}}). \label{RpeFREE2}%
\end{align}
Alternatively, we may write Eq.~(\ref{erreDELTAA}) assuming LoS security,
which means to replace $R_{\text{pe}}^{(k)}$ with the key rate
\begin{equation}
R_{\text{pe,LoS}}^{(k)}=R_{\text{asy,LoS}}^{(k)}(\tau_{\text{LB}},\bar
{n}_{\text{UB}},\bar{n}_{B}^{\text{UB}}).
\end{equation}
The composable key rate in Eq.~(\ref{erreDELTAA}) is $\varepsilon$-secure
against collective Gaussian attacks [cf Eq.~(\ref{epsSECcollective})].

For the heterodyne protocol, we extend the composable key rate of
Eq.~(\ref{compoHETgen}) to the following expression%
\begin{equation}
R_{\text{gen}}^{\text{het}}\geq\frac{np_{\Delta}p_{\text{ec}}}{N}\left(
R_{\text{pe,het}}^{(k)}-\frac{\Delta_{\text{aep}}}{\sqrt{np_{\Delta}}}%
+\frac{\Theta-\Phi_{np_{\Delta}}}{np_{\Delta}}\right)  , \label{hetDELTAgen}%
\end{equation}
where $n$ must account for the $m_{\text{PL}}$ pilots besides the
$m_{\mathrm{et}}$ energy tests, i.e.,
\begin{equation}
n=N-(m+m_{\text{PL}}+m_{\text{et}})=\frac{N-(m+m_{\text{PL}})}%
{1+f_{\mathrm{et}}},
\end{equation}
and $R_{\text{pe,het}}^{(k)}$ is given by Eqs.~(\ref{RpeFREE1})
and~(\ref{RpeFREE2}) for the case of the heterodyne protocol. This rate has
epsilon security $\varepsilon^{\prime}=K_{np_{\Delta}}^{4}\varepsilon/50$
against general attacks, with $\varepsilon$ being the initial security versus
collective attacks (see Sec.~\ref{KR_sec}).

%%%%%%%%%%%%%%%%%%%%%%%%%%%%%%%%%%%%%%%%%%%%%%%%%%%%%%%%%%%%%%%%%%%%%%%%%%%%%%%%%%%%%%%%%%%%%%%%%%%%%%%%%%%%

We perform a numerical investigation assuming the heterodyne protocol with
LLO. This is now implemented in a post-selection fashion in a way to remove
the (non-Gaussian) effect of fading from the distributed data (see above). We
consider the protocol parameters in Table~\ref{tablePARAMETERS} but where we
include the pilots $m_{\text{PL}}=0.05\times N$, so the key generation signals
are reduced to $n\simeq7.08\times10^{6}$, and a threshold parameter
$f_{\text{th}}=0.8$ for post-selection. We then assume the physical parameters
in Table~\ref{WirelessTABLE}, but taking a higher clock value $C=33~$MHz and
also including the transmitter's pointing error $\tilde{\sigma}_{\text{P}}$,
equal to $1/10$ of degree. In this regime of parameters, we study the
composable key rates that are achievable under the various security and trust
assumptions, considering a mobile device which can move up to a maximum
distance $z_{\max}$ from the transmitter (range of the wireless network).

\begin{figure}[t]
\vspace{0.2cm}
\par
\begin{center}
\includegraphics[width=0.9\columnwidth] {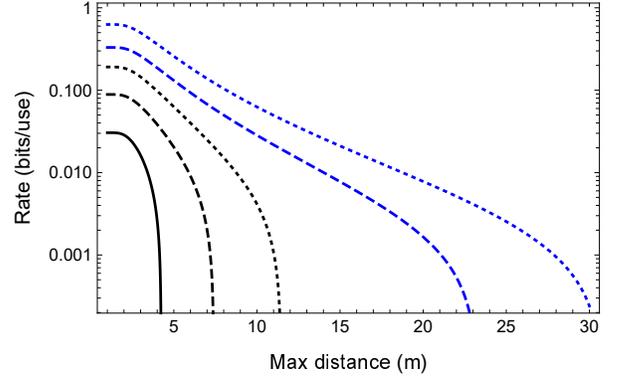}
\end{center}
\par
\vspace{-0.4cm}\caption{Optical-wireless QKD\ with mobile devices. We plot the
composable secret key rate (bits/use) versus the maximum free-space distance
$z_{\max}$ of the receiver-device from the transmitter (meters). This is for a
pilot-guided post-selected heterodyne protocol with an LLO. We show the rates
against collective attacks assuming a trusted-loss-and-noise receiver (black
dotted), a trusted-noise receiver (black dashed), and an untrusted receiver
(solid black). The blue lines refer to line-of-sight security for
trusted-loss-and-noise receiver (blue dotted), and trusted-noise receiver
(blue dashed). Physical parameters are chosen as discussed in the main text.}%
\label{mobilePIC}%
\end{figure}

The rates are plotted in Fig.~\ref{mobilePIC}. Note that the values in the
range of $10^{-2}-1$ bit/use correspond to high rates in the range of
$0.33-33~$Mbits/sec at the considered clock. This means that quantum-encrypted
wireless communication at about $1~$Mbit/sec are possible within distances of
a few meters. Another important consideration is that these rates are actually
lower bounds, since they are computed with the device at the maximum distance
and bounding the noise. This is also the reason why the key rate of
Eq.~(\ref{hetDELTAgen}) does not appear for this specific choice of
parameters. 

\subsection{Short-range microwave wireless}

Let us consider wireless quantum communications at the microwave frequencies,
in particular at 1~GHz. We show the potential feasibility for short-range
quantum-safe WiFi (e.g., for contact-less cards) within the general setting of
composable finite-size security. First of all we need to remark two important
differences with respect to the optical case: presence of higher loss and
higher noise.

From the point of view of increased loss, the crucial difference is the
geometry of the beam. For indoor wireless applications, microwave antennas are
small and, for this reason, cannot offer beam directionality. The emitted beam
is either isotropic (spherical wave) or have some limited directionality,
usually quantified by the gain $g$. This means that, at some distance $z$, the
intensity of the beam will be confined in an area equal to $4\pi z^{2}/g$. It
is clear that we have a strong suppression of the signal, since a receiver
with aperture's radius $a_{R}$ is going to collect just a fraction
$\eta_{\text{ch}}\simeq\min\{ga_{R}^{2}/(4\pi z^{2}),1\}$ of the emitted
photons. Here the minimum accounts for the case where the receiver is close to
the antenna, so the angle of emission is subtended by the receiver's aperture,
which happens at the distance $z_{\text{best}}=\sqrt{g/\pi}a_{R}/2$. In our
investigation, we assume the numerical value $g=10$.

As mentioned above another important difference with respect to the optical
case is the amount of thermal background noise which affects microwaves for
both signal preparation and
detection~\cite{refA1,refA2,refA3,refA4,refA5,refA6}. If we assume setups
working at room temperature, this thermal noise is dominant with respect to
the other sources of noise. Both the preparation noise at the microwave
modulators and the electronic noise in the amplifiers of the microwave
homodyne detectors are relevant~\cite{Shabir}; we set them to be equal to the
thermal background computed using the formula of the black-body radiation. On
the other hand, phase-errors associated with the LO are negligible since the
LO is slow at the microwave and can easily be reconstructed.

Let us quantify the amount of thermal noise and identify a suitable set of
parameters able to mitigate the problem. For a receiver with spectral filter
$\Delta\lambda$, detector bandwidth $W$, aperture $a_{R}$, and field of view
$\Omega_{\text{fov}}$, we can consider the photon collection parameter
$\Gamma_{R}$ in Eq.~(\ref{downT}). Assume that signal and LO pulses are
time-bandwidth limited, so that $\Delta t\Delta\nu\simeq1$. For instance
$\Delta t=10$~ns and $\Delta\nu=100~$MHz for a carrier frequency of $\nu
=1~$GHz ($10\%$ bandwidth). Corresponding carrier wavelength is $\lambda
=c/\nu\simeq30~$cm. Using $\Delta\lambda=\Delta\nu\lambda^{2}/c$ and setting
$W\simeq\Delta\nu$ (detector resolving the pulses), we may write%
\begin{equation}
\Gamma_{R}\simeq\frac{\lambda^{2}}{c}\Omega_{\text{fov}}a_{R}^{2}.
\end{equation}
For receiver aperture $a_{R}=5~$cm and sufficiently-narrow field of view
$\Omega_{\text{fov}}^{1/2}=1~$degree (so $\Omega_{\text{fov}}\simeq
3\times10^{-4}$~sr), we compute $\Gamma_{R}\simeq2.28\times10^{-16}$ in units
of s m$^{3}$ sr. Note that realizing such a narrow field of view with a small
indoor receiver can be challenging in practice.

The photon collection parameter must be combined with the thermal background
photons in units of photons s$^{-1}$ m$^{-3}$ sr$^{-1}$, quantified by the
black-body formula%
\begin{equation}
\bar{n}_{\text{body}}=\frac{2c}{\lambda^{4}}\left[  \exp\left(  \frac
{hc}{\lambda k_{\text{B}}T}\right)  -1\right]  ^{-1},
\end{equation}
where $k_{\text{B}}$ is Boltzmann's constant and $T\simeq290~$K is the
temperature. Therefore we get%
\begin{equation}
\bar{n}_{\text{th}}=\Gamma_{R}\bar{n}_{\text{body}}\simeq0.1\text{~photons.}
\label{nthexp}%
\end{equation}
Note that the figure is acceptably low thanks to the filtering effect of
$\Gamma_{R}$, which accounts for the spatiotemporal profile of the LO pulses,
together with the other features of the receiver (aperture, field of view).

Thermal noise is affecting both preparation and detection with constant floor
level. This means that $\bar{n}_{\text{th}}$ mean photons are seen by the
detector no matter if signal photons are present or not. In other words, the
detector experiences a constant noise variance equal to
\begin{equation}
\sigma_{z}^{2}=2\bar{n}_{\text{th}}+\nu_{\text{det}}, \label{noiseZmicro}%
\end{equation}
where $\nu_{\text{det}}$ is the usual quantum duty (which is $=1$ for homodyne
and $=2$ for heterodyne).

Assume that the total transmissivity is $\tau=\eta_{\text{ch}}\eta
_{\text{eff}}$, where $\eta_{\text{ch}}$ is channel's transmissivity and
$\eta_{\text{eff}}\simeq0.8$ is receiver's efficiency. Also assume that the
transmitter (Alice), modulates thermal states with classical variance
$\sigma_{x}^{2}=2\bar{n}_{T}$, where $\bar{n}_{T}$ is equivalent mean number
of signal photons. Then, the total mean number of photons at the receiver's
detector is given by%
\begin{equation}
\bar{n}_{R}=\tau\bar{n}_{T}+\bar{n}_{\text{th}}. \label{IOmicro}%
\end{equation}
Basically, this is equivalent to Eqs.~(\ref{IOenergy}) and~(\ref{nBARvalue}),
by setting $\bar{n}_{B}=\bar{n}_{\text{th}}$ and $\bar{n}_{\text{ex}}%
=(1-\eta_{\text{eff}})\bar{n}_{\text{th}}$. As we can see, for $\tau=1$, we
get $\bar{n}_{T}+\bar{n}_{\text{th}}$ meaning that the prepared states are
thermal; for $\tau<1$, signal photons are lost ($\bar{n}_{T}\rightarrow
\tau\bar{n}_{T}$), while the depleted thermal background photons are
compensated at the receiver re-entering the detection system, so we have the
constant noise level $\bar{n}_{\text{th}}$.

\subsubsection{Fully-untrusted scenario}

In the worst-case scenario, the noise associated with preparation, channel and
detector is all untrusted. In this case, Eq.~(\ref{IOmicro}) corresponds to
the action of a beam splitter with transmissivity $\tau$ combining a signal
mode with mean photons $\bar{n}_{T}$ and an environmental mode with mean
photons $\bar{n}_{e}=\bar{n}_{\text{th}}/(1-\tau)$. The idea is that Alice
would attempt to create randomly-displaced coherent states, but Eve readily
thermalizes them by adding malicious thermal photons. These photons add up to
those later introduced by the channel, so that we globally have the insertion
of $\bar{n}_{e}$ mean photons as above. This leads to a collective Gaussian
attack where Eve has the purification of the untrusted thermal noise
associated with each stage of the communication.

Alice's and Bob's classical variables, $x$ and $y$, are related by
Eq.~(\ref{IOnew}) but where the noise variable $z$ has now variance
$\sigma_{z}^{2}$ as in Eq.~(\ref{noiseZmicro}) which corresponds to
Eq.~(\ref{sigmazed}) up to replacing $\bar{n}\rightarrow\bar{n}_{\text{th}}$.
Alice and Bob's mutual information $I(x:y)$ is therefore given by
Eq.~(\ref{MutualINFOeq}) computed with modulation $\sigma_{x}^{2}=2\bar{n}%
_{T}$ and equivalent noise%
\begin{equation}
\chi=\frac{2\bar{n}_{\text{th}}+\nu_{\text{det}}}{\tau}=\xi_{\text{tot}}%
+\frac{\nu_{\text{det}}}{\tau},
\end{equation}
where $\xi_{\text{tot}}:=2\bar{n}_{\text{th}}/\tau$ is the total excess noise.
Numerically, we choose the modulation $\sigma_{x}^{2}=20$.

As already said, in the fully-untrusted scenario, all thermal noise coming
from preparation, channel and receiver's setup is considered to be untrusted.
This is equivalent to the treatment of Sec.~\ref{AsyUntrustedSection}, proviso
we make the replacement $\bar{n}\rightarrow\bar{n}_{\text{th}}$ in
Eq.~(\ref{untrustedOMEGA}) and then in Eqs.~(\ref{parameterB}),
(\ref{thetaEQ1}) and~(\ref{fiEQ1}). The revised parameters can then be used in
the global CM\ in Eqs.~(\ref{jointCM}) and~(\ref{CMcase1}).

Then, the asymptotic key rate against collective Gaussian attacks is given by
$R_{\text{asy}}^{(3)}(\tau,\bar{n}_{\text{th}})$ according to
Eq.~(\ref{asyUNrate}), where we now use%
\begin{equation}
\tau=\eta_{\text{eff}}\min\{ga_{R}^{2}/(4\pi z^{2}),1\}, \label{tauMICRO}%
\end{equation}
and $\bar{n}_{\text{th}}$ as given by Eq.~(\ref{nthexp}). We may then assume
the reconciliation parameter $\beta=0.98$.

To account for finite-size effects, we first include parameter estimation. This
means that the parties need to sacrifice $m$ of the $N$\ pulses, so $n$ pulses
survive for key generation. Numerically, we take $N=5\times10^{7}$ and
$m=0.1\times N$. Thus, they construct the worst-case estimators for the
overall transmissivity $\tau$ and thermal noise $\bar{n}_{\text{th}}$
following Eqs.~(\ref{wcEstimator1}) and~(\ref{wcEstimator2}). These estimators
can be here approximated as follows%
\begin{align}
\tau^{\prime}  &  \simeq\tau-2w\sqrt{\frac{2\tau^{2}+\tau(2\bar{n}_{\text{th}%
}+\nu_{\text{det}})/\sigma_{x}^{2}}{\nu_{\text{det}}m}},\label{estREV1}\\
\bar{n}_{\text{th}}^{\prime}  &  \simeq\bar{n}_{\text{th}}+w\frac{2\bar
{n}_{\text{th}}+\nu_{\text{det}}}{\sqrt{2\nu_{\text{det}}m}}, \label{estREV2}%
\end{align}
where $w$ is the confidence parameter associated with $\varepsilon_{\text{pe}%
}$, and computed according to Eq.~(\ref{wSTVALUE}) for collective Gaussian
attacks (see Sec.~\ref{PEsectionLAter} for more details). Assuming a tolerable
error probability of $\varepsilon_{\text{pe}}=2^{-33}$, we have $w\simeq6.34$
confidence intervals.

The composable key rate takes the form in Eq.~(\ref{sckeee}) where we now use
$R_{\text{pe}}^{(3)}=R_{\text{asy}}^{(3)}(\tau^{\prime},\bar{n}_{\text{th}%
}^{\prime})$ computed from Eqs.~(\ref{estREV1}) and~(\ref{estREV2}), together
with the usual finite-size terms in Eqs.~(\ref{deltaAEPPP})
and~(\ref{bigOMEGA}). Numerically, we can assume $p_{\text{ec}}=0.9$ for the
probability of success of EC, $d=2^{5}$ for the digitalization of the
continuous variables, and the value $2^{-33}$ for all the epsilon parameters,
so we have epsilon security $\varepsilon\simeq5.6\times10^{-10}$ against
collective Gaussian attacks according to Eq.~(\ref{epsSECcollective}).

To study the performance, let us consider the heterodyne protocol
($\nu_{\text{det}}=2$). Then, we assume a device stably kept at some distance
$z$ from the transmitter within the emission angle of the transmitter and with
an aligned field of view. For the parameters considered here, we find that a
positive key rate is obtained for $z\leq4.48$~cm, which is fully compatible
for contactless card applications. In particular, for any $z\leq
z_{\text{best}}\simeq4.46$~cm we compute a key rate of $R\gtrsim10^{-2}$
bits/use, corresponding to $\gtrsim50$ kbit/sec with a system clock at $5$ MHz.

Note that, according to the thermal version of the PLOB bound~\cite{QKDpaper},
the maximum key rate cannot overcome the upper limit
\begin{equation}
R\leq\left\{
\begin{array}
[c]{l}%
-\log_{2}\left[  (1-\tau)\tau^{\frac{\bar{n}_{\text{th}}}{1-\tau}}\right]
-h\left(  \frac{\bar{n}_{\text{th}}}{1-\tau}\right)  ,~~\text{for~}\bar
{n}_{\text{th}}\leq\tau,\\
0,~~\text{for~}\bar{n}_{\text{th}}\geq\tau,
\end{array}
\right.
\end{equation}
where $h(x):=H(2x+1)$. This means that the no rate is possible above the
threshold $\bar{n}_{\text{th}}=\tau$. Using Eqs.~(\ref{nthexp})
and~(\ref{tauMICRO}) with our regime of parameters, we find that the maximum
possible range is about $12.47~$cm, i.e., about three times the distance
achievable with the considered heterodyne protocol under composable security.

\subsubsection{LoS security for microwaves}

Better performances can be obtained if we relax security requirements by
relying on the LoS geometry. In particular, one may assume that the thermal
noise is trusted, so that Eve is passively limited to eavesdrop the photons
leaking from the channel and the setup. In this case, Eq.~(\ref{IOmicro})
corresponds to the action of a beam splitter with transmissivity $\tau$
combining a signal mode with mean photons $\bar{n}_{T}+\bar{n}_{\text{th}}$
(signal photons plus trusted preparation noise)\ and a genuine environmental
mode with mean photons $\bar{n}_{\text{th}}$~\cite{NotaNOISE}. Eve collects
the fraction $1-\tau$ of photons leaked into the environment, but she does not
control any noise, i.e., she does not have its purification.

Alice and Bob's mutual information $I(x:y)$ is the same as above for the
fully-untrusted case but Eve's Holevo information $\chi_{\text{LoS}}(E:y)$ is
now rather different. The latter can be computed as in Sec.~\ref{LoSsection}
and, in particular, from the CM\ in Eq.~(\ref{VbeLOS}), where we insert the
following parameters%
\begin{align}
b  &  =2\bar{n}_{R}+1,\\
\theta &  =-\sqrt{\tau(1-\tau)}\sigma_{x}^{2},\\
\phi &  =(1-\tau)\sigma_{x}^{2}+2\bar{n}_{\text{th}}+1.
\end{align}
In this way we can compute the asymptotic key rate%
\begin{equation}
R_{\text{asy,LoS}}(\tau,\bar{n}_{\text{th}})=\beta I(x:y)-\chi_{\text{LoS}%
}(E:y).
\end{equation}

The incorporation of finite-size effects requires that we under-estimate the
thermal noise experienced by Eve, while we over-estimate that seen by the
parties. Thus, besides the worst-case estimators $\tau^{\prime}$ and $\bar
{n}_{\text{th}}^{\prime}$ in Eqs.~(\ref{estREV1}) and~(\ref{estREV2}), we also
compute the best-case estimator%
\begin{equation}
\bar{n}_{\text{th}}^{\prime\prime}\simeq\bar{n}_{\text{th}}-w\frac{2\bar
{n}_{\text{th}}+\nu_{\text{det}}}{\sqrt{2\nu_{\text{det}}m}}.
\end{equation}
Thus, we compute the rate
\begin{equation}
R_{\text{pe,LoS}}=\beta I(x:y)_{\tau^{\prime},\bar{n}_{\text{th}}^{\prime}%
}-\chi_{\text{LoS}}(E:y)_{\tau^{\prime},\bar{n}_{\text{th}}^{\prime\prime}},
\end{equation}
which is replaced into Eq.~(\ref{sckeee}) to provide the composable key rate
associated with LoS security.

\begin{figure}[t]
\vspace{0.2cm}
\par
\begin{center}
\includegraphics[width=0.9\columnwidth] {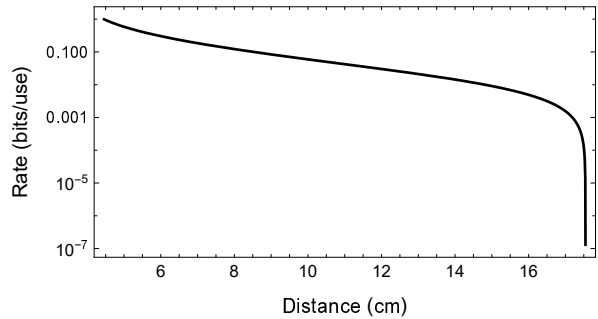}
\end{center}
\par
\vspace{-0.4cm}\caption{Microwave wireless QKD (at 1~GHz) using the heterodyne
protocol under LoS security. We plot the composable secret key rate (bits/use)
versus free-space distance $z$ between transmitter and receiver (centimeters).
Parameters are chosen as discussed in the main text.}%
\label{WiFipic}%
\end{figure}

Assuming the heterodyne protocol with the same parameters as in the
fully-untrusted case, we find an improvement, as expected. As shown in
Fig.~\ref{WiFipic}, the range of security is now larger, even though the
effective application is still restricted to centimeters from the transmitter.
Note that this performance is based on the LoS assumption, so it is not
confined by the PLOB bound.

\section{Conclusions\label{SEC3}}

In this work, we have developed a general framework for the composable
finite-size security analysis of Gaussian-modulated coherent-state protocols,
which are the most powerful protocols of CV-QKD. We have investigated the
secret key rates that are achievable assuming various levels of trust for the
receiver's setup, from the worst-case assumption of a fully-untrusted detector
to the case where detector's loss and noise are considered to be trusted. In
the specific case of free-space quantum communication, we have also
investigated the additional assumption of passive eavesdropping on the
communication channel due to the line-of-sight geometry.

We have shown how the realistic assumptions on the setups can have non-trivial
effects in terms of increasing the composable key rate and tolerating higher
loss (therefore increasing distance). More interestingly, we have also
demonstrated the feasibility of high-rate CV-QKD with wireless mobile devices,
assuming realistic parameters and near-range distances, e.g., as typical of
indoor networks. Besides the optical frequencies, we have also analyzed the
microwave wavelengths, considering possible parameters able to mitigate the
loss and noise affecting this challenging setting. In this way, we have
discussed potential microwave-based applications for very short-range
(cm-range) quantum-safe communications.

\bigskip

\textit{Acknowledgements}.~~The author would like to thank Panagiotis
Papanastasiou, Masoud Ghalaii, Cillian Harney, and Marco Tomamichel for
discussions. This work was funded by the European Union's Horizon 2020
research and innovation programme under grant agreement No 820466
(Quantum-Flagship Project CiViQ: \textquotedblleft Continuous Variable Quantum
Communications\textquotedblright).

%\textit{Acknowledgements}.--~The author acknowledges funding from the European
%Union's Horizon 2020 research and innovation programme under grant agreement
%No 820466 (Quantum-Flagship Project CiViQ: \textquotedblleft Continuous
%Variable Quantum Communications\textquotedblright).


\begin{thebibliography}{99}
\bibitem {noteDalpha}On the other hand, discrete-alphabet CV-QKD protocols
based on coherent-state constellations have more limited security proofs,
asymptotic versus general attacks~\cite{FA0,FA1}, with very recent composable
analyses~\cite{PanosFA,Koashi}.
\bibitem {NoteEstimator}The approximations are valid up to $O(m_{p}^{-1})$.
Note that the expression of $\mathrm{var}(\hat{\tau})$ may be further
approximated for large $m_{p}$, so one can write the more optimistic estimator
$\tau^{\prime}\simeq\tau-2w\sigma_{z}/\sigma_{x}\sqrt{\tau/m_{p}}$.
\bibitem {NotePilots}Because parameter estimation will be performed over the
signals, Alice and Bob are able to detect eavesdropping strategies where Eve's
interaction is different between signals and pilots (e.g., assuming that Eve
had the capability to discriminate between these two types of signals via a
quantum non-demolition measurement). In the presence of such asymmetric
strategies, the parties are able to estimate the corresponding deformation of
the post-selection interval and to account for extra signal loss in the
post-processing and final key rate. See Ref.~\cite{FSpaper}\ for more details.
\bibitem {NotaNOISE}Note that this means that the environment has noise
variance $\tilde{\omega}=2\bar{n}_{\text{th}}+1$, and the variance of Alice's
average state is $\tilde{\mu}=\sigma_{x}^{2}+\tilde{\omega}$, i.e., classical
modulation plus trusted thermal noise.
\end{thebibliography}
\end{document}